\documentclass[aps,prb,amsmath,amssymb, superscriptaddress, preprint]{revtex4-1}
\usepackage[utf8]{inputenc} % set input encoding (not needed with XeLaTeX)

\usepackage{geometry} % to change the page dimensions
\usepackage{graphicx} % support the \includegraphics command and options
\usepackage{color}
\usepackage{booktabs} % for much better looking tables
\usepackage{array} % for better arrays (eg matrices) in maths
\usepackage{paralist} % very flexible & customisable lists (eg. enumerate/itemize, etc.)
\usepackage{verbatim} % adds environment for commenting out blocks of text & for better verbatim
\usepackage{subfig} % make it possible to include more than one captioned figure/table in a single float
\usepackage{hyperref}
\newcommand*{\citen}[1]{%
  \begingroup
    \romannumeral-`\x % remove space at the beginning of \setcitestyle
    \setcitestyle{numbers}%
    \cite{#1}%
  \endgroup   
}

\makeatletter
\newcommand*{\rom}[1]{\expandafter\@slowromancap\romannumeral #1@}
\makeatother

\def\bra#1{\langle#1| }
\def\ket#1{| #1 \rangle }

\begin{document}

\title{A universal approach to quantum thermodynamics in the strong coupling regime}
%\author{Wenjie Dou$^{1}$, Maicol A. Ochoa$^{1*}$, Abraham Nitzan$^{1,2}$, and Joseph E. Subotnik$^{1}$}
%\affiliation{$^{1}$ Department of Chemistry, University of Pennsylvania, Philadelphia, Pennsylvania 19104, USA  \\
%$^{2}$School of Chemistry, The Sackler Faculty of Science, Tel Aviv University, Tel Aviv 69978, Israel}

\author{Wenjie Dou}
\affiliation{Department of Chemistry, University of Pennsylvania, Philadelphia, Pennsylvania 19104, USA}
\author{Maicol A. Ochoa}
\thanks{Present address: 
Center for Nanoscale Science and Technology, National Institute of Standards and Technology, Gaithersburg, MD 20899 
\& 
Institute for Research in Electronics and Applied Physics, University of Maryland, College Park, MD 20742.}
\affiliation{Department of Chemistry, University of Pennsylvania, Philadelphia, Pennsylvania 19104, USA}
\author{Abraham Nitzan}
\affiliation{Department of Chemistry, University of Pennsylvania, Philadelphia, Pennsylvania 19104, USA}
\affiliation{School of Chemistry, The Sackler Faculty of Science, Tel Aviv University, Tel Aviv 69978, Israel}
\author{Joseph E. Subotnik}
\affiliation{Department of Chemistry, University of Pennsylvania, Philadelphia, Pennsylvania 19104, USA}

\begin{abstract}

We present a protocol for the study of the dynamics and thermodynamics of quantum systems strongly coupled to a bath and subject to an external modulation. Our protocol quantifies the evolution of the system-bath composite by expanding the full density matrix as a series in the powers of the modulation rate, from which the functional form of work, heat and entropy rates can be obtained. Under slow driving, thermodynamic laws are established. The entropy production rate is positive and is found to be related to the excess work dissipated by friction, at least up to second order in the driving speed. As an example of the present methodology, we reproduce the results for the quantum thermodynamics of the driven resonance level model. We also emphasize that our formalism is quite general and allows for electron-electron interactions, which can give rise to exotic Kondo resonances appearing in thermodynamic quantities.

\end{abstract}

\maketitle
%%%%%%%%%%%%%%%%%%%%%%%%%%%%%%%%%%%%%%%%%%%%%%%%%%%%%%%

\section{Introduction}

Modern nanofabrication techniques, super-resolution spectroscopies and nanoscale sensors provide tools for the design, control and study of systems made up of just a few atoms, namely, far from the thermodynamic limit. In this regime, both thermal and quantum mechanical fluctuations are essential and cannot be neglected. Biomolecular motors, driven transport {nanojunctions} and quantum computing elements stand out as prototypical {  nanoscale systems that can be utilized to perform tasks} under continuous energy exchange with their surroundings. Just as for macroscopic engines, in order to understand the nature of the work {  performed} and heat produced at this scale, both theoretical principles and computational methods for evaluating energy conversion and thermodynamic efficiency are necessary.  From a conceptual point of view, the modern field of quantum thermodynamics\cite{gemmer2009quantum,kosloff2013quantum, PhysRevLett.114.080602, vinjanampathy2016quantum,anders2017focus,PhysRevLett.116.240403,alicki2018introduction,benenti2017fundamental,RevModPhys.83.771,PhysRevLett.102.210401} addresses these questions while accounting for the quantum nature of the nanoscopic system. As a result, concepts such as quantum dissipation and frictional effects have been formulated in thermodynamic terms. \cite{mishaPRBfriction,millen2016perspective, gemmer2009quantum, talkner2007fluctuation,brandao2015second,jarzynski2011equalities} Recent experimental works\cite{poot2012mechanical, pekola2015towards, rossnagel2016single,PhysRevE.96.052106} have started to address these concepts.

Under the condition of weak coupling strength between a system and its surroundings, quantum thermodynamics can successfully describe \cite{spohn1978irreversible,RevModPhys.81.1665,kosloff2013quantum,gelbwaser2015thermodynamics,PhysRevLett.119.050601} the dynamics, heat, work and entropy production rates for the system-bath composite in terms of the reduced density matrix for the system. By contrast, the strong coupling regime has proven more challenging, and advanced strategies have been necessary.\cite{PhysRevB.98.085415} For example, strongly coupled quantum heat engines \cite{gelbwaser2015strongly} {  have} been investigated with a polaron transformation, which can somehow map the original strongly coupled system into an equivalent weakly coupled system. Another approach is to introduce heat exchangers \cite{uzdin2016quantum,katz2016quantum}, essentially increasing the system space to accommodate strongly coupled environmental modes. The concept of active and passive states \cite{pusz1978passive,lenard1978thermodynamical,janzing2006computational,niedenzu2016operation}, which identifies unitary transformations acting on the space of reduced density matrices for the system with processes that can potentially deliver work, {  provides} yet another approach. Lastly, techniques from quantum information theory can also provide some quantum thermodynamic principles\cite{parrondo2015thermodynamics, pekola2016maxwell, averin2017reversing}. %The resonance level model \cite{PhysRevLett.114.080602,PhysRevLett.119.050601, PhysRevB.93.115318,PhysRevB.94.035420,haughian2018quantum,PhysRevB.89.161306,bruch2018landauer} has recently attracted
%attention as a simple platform for studying quantum thermodynamics of systems that are strongly
%coupled to their environment.

In general, for  the strong coupling regime, the key point to emphasize is that the reduced density matrix of the embedded system does not necessarily contain all the information needed to describe the dynamics and thermodynamics of that same system; instead, one needs to include corrections originating from the system-bath {  couplings}. Several recent works, e.g. Refs.\ \citenum{solinas2013work,schmidt2015work,gogolin2016equilibration,subacsi2012equilibrium,ness2017nonequilibrium,PhysRevX.7.021003}, have
addressed this situation. For instance, in Ref. \citenum{solinas2013work} work is defined in terms of the reduced density matrix and the total power dissipated during the evolution of the full composite system (i.e. system+bath). This approach was later refined \cite{schmidt2015work} to include the action of strong external fields within the stochastic Liouville-von Neumann scheme, capturing the full non-Markovian {  nature of} the reduced density matrix. In both cases, the initial state is taken to be a tensor product of the quasithermal state for the system and the thermal state for the bath, which can be used when the focus is on the long time steady state or periodic behavior \cite{gogolin2016equilibration,subacsi2012equilibrium}. When applied to open systems at {  steady state}, such an approach permits the computation of the entropy production for the system in the presence of {  two reservoirs with different temperatures and chemical potentials}, in both the weak and the strong coupling  regimes \cite{ness2017nonequilibrium}.  Recently, an approach based on effective quantum master equations   has been formulated, relying on a protocol by which the system repeatedly interacts with identically prepared ``units" \cite{PhysRevX.7.021003}. The overall consequences of these interactions can be assessed from the initial and final state of these units, which  results in a propagation scheme for the system density matrix that is consistent with the correct thermodynamics in a few model systems. A conceptually similar setup, describing {  thermodynamic processes} as a sequence of quenches and thermalization processes affected by turning off and on system-baths interactions, {  has been} recently explored by Perarnau-Llobet and co-workers \cite{PhysRevLett.120.120602}.

From our perspective, 
{  a generic and universal approach to quantum thermodynamics that interpolates between weak and strong coupling regimes is still lacking even in  classical mechanics \cite{PhysRevLett.116.020601, talkner2016open,PhysRevX.7.011008}, due to the inherent difficulties of defining proper thermodynamic quantities (e.g. internal energy, heat, work and entropy) consistent with the thermodynamic laws.} Likewise, the definition of thermal properties of the system such as heat capacity\cite{hanggi2008finite,ingold2009specific} may lead to apparent anomalies.  A prototypical model system that has served as the playground to test new ideas is the driven resonant-level model  \cite{PhysRevLett.114.080602,PhysRevLett.119.050601, PhysRevB.93.115318,PhysRevB.94.035420,haughian2018quantum,PhysRevB.89.161306,bruch2018landauer}. Under the wide-band approximation, with proper splitting of the system-bath coupling, a consistent thermodynamic description with a proper formulation of the first and second laws can be established for the averaged thermodynamic observables, including situations where the system-bath coupling is time dependent  \cite{haughian2018quantum}. However, the coupling splitting assumption may fail to reproduce higher moments in the energy distribution\cite{PhysRevB.94.035420}. 

With this background in mind, 
 in the present paper, we will study the quantum thermodynamics of general systems from the perspective of the full density matrix for the system-bath composite. The {  present method} does not make any assumptions regarding the complexity of the original system or  bath, {  allowing the inclusion of interactions and treating fermionic and bosonic systems} on an equal footing. {  This scheme} can be seen as a natural extension of the strategies developed in Refs.\ \citenum{PhysRevB.93.115318} and \citenum{ochoa2018quantum} for strongly coupled systems near equilibrium. Our key ingredient for defining consistent thermodynamic quantities relies on separating the dynamic evolution of the full density operator of the driven system into the explicit time evolution and the (assumed slow) driving terms \cite{PhysRevLett.119.050601, PhysRevLett.119.046001}:
\begin{eqnarray} \label{eq:first}
\frac{d}{dt} = \frac{\partial }{\partial t} + \sum_\alpha \dot R^{\alpha} \partial_\alpha ,
\end{eqnarray}
where $R^{\alpha}$ are system parameters that are modulated over time, with $\dot R^{\alpha}$ being the imposed driving speeds. Stating from the equation of motion for the full density operator and assuming different timescales for the internal time evolution $\frac{\partial }{\partial t}$ and the driving processes  $\dot R^{\alpha}$, we obtain the full density operator in a power series in the driving speeds beyond first order\cite{beilstein}. 

By implementing this approach we will be able to consistently define thermodynamic quantities of the full, composite system as well as for the driven sub-system, which naturally reduce to their equilibrium values at vanishing driving speeds. When applied to the driven resonant level model, this approach matches previously obtained previous results.\cite{PhysRevLett.114.080602, PhysRevB.93.115318,PhysRevB.94.035420} Moreover, beyond most approaches, our formalism allows for electron-electron interactions, which we shall demonstrate gives rise to interesting Kondo resonances in the evaluated thermodynamic quantities. Up to the second order in the driving rate, we find that entropy production is related to frictional work. For the case of one fermionic or bosonic bath, such friction is positive definite, in agreement with the second law of thermodynamics. However, as will become clear below, the proper definition of entropy above second order in the driving rates and in presence of multiple baths under nonequilibrium conditions will require further studies.

To avoid confusion, a note should be made about language. The system of interest is a driven microscopic sub-system that interacts, possibly strongly, with its macroscopic environment. Together, this subsystem and its environment constitute a macroscopic system that we refer to as the total, full, or composite system. This full system can be treated within macroscopic thermodynamics as a closed system or as a system open to energy exchange (canonical) or energy and particle exchange (grand canonical) with an even larger equilibrium environment (referred below as ``superbath") characterized by temperature $T$ and chemical potential $\mu$. One may safely assume that the dynamics and thermodynamic properties (assuming the latter can be defined) of the driven microscopic system do not depend on the nature of the interaction between the ``full system" and the ``superbath", however we will see that some thermodynamic considerations may depend on how this interaction is taken into account.

We organize the paper as follows. In Sec. \ref{sec:static}, we establish the first law of thermodynamics in the quasi-static limit. In Sec. \ref{sec:nonadiabatic}, we extend the results into the finite speed, and connect the entropy change to frictional work.  In Sec. \ref{sec:sub-system}, we introduce system-bath separation, reformulate thermodynamics law for the sub-system, and apply the results to the resonant-level model. An interesting Kondo resonance in thermodynamic quantities shows up when electron-electron interactions are included. We conclude in Sec. \ref{sec:con}.

\section{Static and quasi-static thermodynamics}  \label{sec:static}

We start by reviewing the quasi-static (reversible) limit. In this limit, the modulation of Hamiltonian parameters is done slowly enough relative to relaxation processes that bring the system to equilibrium. Consequently, the system evolves adiabatically while remaining at equilibrium with its environment for the instantaneous values of the modulated parameters.

\subsection{Static thermodynamics}
We consider a very general large system consisting fermions and/or bosons (or both) at equilibrium with an environment characterized by a temperature $T=k_B \beta^{-1}$ and chemical potential $\mu$. The equilibrium
density operator $\hat \rho^{(0)}$ is
\begin{eqnarray} \label{eq:rho0}
\hat \rho^{(0)} = e^{-\beta (\hat H-\mu \hat N )}/\Omega,
\end{eqnarray}
where $\Omega$ is the grand canonical partition function, $\Omega = {\rm Tr} (e^{-\beta ( \hat H-\mu \hat N )})$, and  $\hat N$ is the particle number operator. We consider the case where $\hat H$ commutes with $\hat N$, such that $\hat \rho^{(0)}$ is well defined. \footnote{This implies that we disregard the coupling responsible for the exchange of particles between system and environment as is the practice for describing equilibrium systems in the thermodynamic limit.} The static free energy (grand potential) $F^{(0)}$, defined by
\begin{eqnarray} \label{eq:F}
F^{(0)} = - \frac 1\beta  \ln \Omega . 
\end{eqnarray}
can be used to calculate other thermodynamic quantities, such as the static number of particles $N^{(0)}$
\begin{eqnarray} \label{eq:N}
N^{(0)} = -\frac{\partial}{\partial \mu}  F^{(0)}   =   {\rm Tr}( \hat N \hat \rho^{(0)}), 
\end{eqnarray}
and the static entropy $S^{(0)}$ 
\begin{eqnarray} \label{eq:Si}
S^{(0)} &=&-\frac{\partial F^{(0)}}{\partial T}= k_B \beta^2 \frac{\partial}{\partial \beta} F^{(0)} = k_B \left( \ln \Omega  + \beta  {\rm Tr}( \hat H \hat \rho^{(0)}) -  \beta \mu {\rm Tr}( \hat N \hat \rho^{(0)} )  \right) .
\end{eqnarray}
so that the static system energy
\begin{eqnarray} \label{eq:E}
E^{(0)} =  {\rm Tr}( \hat  H \hat  \rho^{(0)}), 
\end{eqnarray}
satisfies
\begin{eqnarray} \label{eq:law0}
E^{(0)} = F^{(0)} + T S^{(0)} + \mu N^{(0)} .
\end{eqnarray}
Finally, from the definition of $\hat \rho^{(0)}$ and using $ {\rm Tr} \hat  \rho^{(0)} =1$, we find 
\begin{eqnarray}
 {\rm Tr}(\hat \rho^{(0)} \ln \hat \rho^{(0)} ) 
 = - \ln \Omega - \beta {\rm Tr}( \hat  H \hat  \rho^{(0)}) +  \beta \mu {\rm Tr}( \hat  N \hat  \rho^{(0)} ), 
\end{eqnarray}
such that  the static entropy $S^{(0)}$, Eq. \eqref{eq:Si}, can be rewritten as 
\begin{eqnarray} \label{eq:S}
S^{(0)} =  - k_B  {\rm Tr}(\hat  \rho^{(0)} \ln \hat  \rho^{(0)} ) . 
\end{eqnarray}

%%%%%%%%%%%%%%%%%%%%%%%%%%%%%%%%%%%%%%%%%%%%%%%%
\subsection{Quasi-static thermodynamics: first order in driving speed}
Now assume that the total system is subject to infinitesimally slow driving, such that $\hat H$ is time dependent. To be specific, let $\hat H$ depends on a parameter set $\bold R = (R^1, R^2, ..., R^\alpha, ... )$, which changes slowly over time.  Henceforward,  the speeds $\{ \dot R^{\alpha} \}$ will be our essential constant parameters. In the quasi-static limit, i.e., %
infinitesimally slow driving (i.e. $\dot{R}^{\alpha} \approx 0$), we assume that the system remains equilibrated at each time step. 
In this limit, the rate of change for all thermodynamic quantities  is simply given by the adiabatic derivative with respect to time, $\frac{d}{dt} = \sum_\alpha \dot R^{\alpha} \partial_\alpha$, of all the equations in the previous subsection. For example, the rate of change of the total energy in the quasi-static limit is, from Eq. \eqref{eq:E}, 
\begin{eqnarray} \label{eq:E1}
\dot{E}^{(1)} = \sum_\alpha \dot{R^{\alpha}} \partial_\alpha E^{(0)} = \sum_\alpha \dot{R^{\alpha}} {\rm Tr}( \partial_\alpha \hat  H \hat  \rho^{(0)}) +  \sum_\alpha \dot{R^{\alpha}} {\rm Tr}(\hat  H \partial_\alpha  \hat  \rho^{(0)}) 
\end{eqnarray}
(The superscript $^{(1)}$ indicates quantities linear in the driving speeds $\{ \dot R^{\alpha} \}$.)  
Similarly, the rate of work done and heat exchanged in this limit are (using Eq. \eqref{eq:F} and Eq. \eqref{eq:Si}, respectively)
\begin{eqnarray} \label{eq:W1}
\dot{W}^{(1)} = \sum_\alpha \dot{R^{\alpha}} \partial_\alpha F^{(0)} = \sum_\alpha \dot{R^{\alpha}} {\rm Tr}( \partial_\alpha \hat  H \hat  \rho^{(0)}), 
\end{eqnarray}
\begin{eqnarray} \label{eq:Q1}
\dot{Q}^{(1)} = T \sum_\alpha \dot{R^{\alpha}} \partial_\alpha S^{(0)} = \sum_\alpha \dot{R^{\alpha}} {\rm Tr}( \hat  H \partial_\alpha \hat  \rho^{(0)}) - \mu \sum_\alpha \dot{R^{\alpha}} {\rm Tr}( \hat  N  \partial_\alpha  \hat  \rho^{(0)}) .
\end{eqnarray}
Finally, from Eq. \eqref{eq:N}, the rate of change of the particle number in the quasi-static limit is 
\begin{eqnarray} \label{eq:N1}
 \dot{N}^{(1)} =  \sum_\alpha \dot{R^{\alpha}} \partial_\alpha N^{(0)} = \sum_\alpha \dot{R^{\alpha}} {\rm Tr}(\hat  N  \partial_\alpha \hat  \rho^{(0)}) .
\end{eqnarray}

Eqs. \eqref{eq:E1}-\eqref{eq:N1} imply that the first law of  thermodynamics  is obeyed in the quasi-static limit, namely, to the first order in $\dot{R}^{\alpha}$: 
\begin{eqnarray} \label{eq:firstLaw1}
\dot{E}^{(1)} = \dot{W}^{(1)} + \dot{Q}^{(1)} + \mu \dot{N}^{(1)} .
\end{eqnarray}

Before leaving this section, a note of caution about the calculation of $\dot Q$ should be made. A popular definition of work and heat in open systems is given in terms of contribution to the total energy change
\begin{eqnarray} \label{eq:dE_dtful}
\frac{dE}{dt} = \frac{d}{dt} {\rm Tr}( \hat  H \hat  \rho ) =    {\rm Tr}( \hat  H \frac{d \hat  \rho }{dt} ) 
+   {\rm Tr}( \frac{d  \hat  H }{dt} \hat  \rho ) =   {\rm Tr}( \hat  H \frac{d \hat  \rho }{dt} )  + \sum_\alpha \dot{R^{\alpha}} {\rm Tr}( \partial_\alpha \hat  H  \hat  \rho)   
\end{eqnarray}
where the second term on the right hand side (RHS) is the work performed on the system and the first term represents the heat that enters (when positive) it, per unit time. As is set now (before we consider thermodynamic functions of subsystems in Sec. \ref{sec:sub-system}), the time evolution under consideration is that of the full system. However, in the closed full system comprising the subsystem of interest and its environment, the so-defined heat vanishes
\begin{eqnarray} \label{eq:dQ/dtful}
\frac{d Q }{dt}  = {\rm Tr}( \hat H  \frac{d \hat \rho }{dt}  ) =  - \frac{i}{\hbar} {\rm Tr}( \hat H [\hat H, \hat \rho] ) = 0 
\end{eqnarray}

Note that the same is true also for the corresponding grand canonical expression $\frac{dQ}{dt}= {\rm Tr}( (\hat H - \mu \hat N)  \frac{d \hat \rho }{dt}  )$ that takes into account possible change in number of particles vanishes for a closed system. Indeed, the existence of finite heat current stems from the recognition that the actual time evolution of the density operator is given by
\begin{eqnarray} \label{eq:Xrho}
\frac{d  \hat \rho }{dt}  =  - \frac{i}{\hbar}  [\hat H, \hat \rho]  - \hat{\hat{\mathcal{X}}} \hat \rho
\end{eqnarray}
where $\hat{\hat{\mathcal{X}}} \hat \rho$ expresses the relaxation dynamics associated with the (small) coupling of the full system to a ``superbath" of temperature $T$ and chemical potential $\mu$. 
This coupling brings the
system to equilibrium for any given constant $\bold R$. However, the coupling is assumed small enough so as not to affect the response of the system to the driving at any finite time. 

%%%%%%%%%%%%%%%%%%%%%%%%%%%%%%%%%%%%%%%%%%%%%%%%
\section{nonadiabatic thermodynamics and entropy production} \label{sec:nonadiabatic}

Having provided the relevant background above, we next go beyond the quasi-static limit and address the case where the system is subject to a finite speed driving. The driven system state now deviates from equilibrium, giving rise to dissipation and entropy production.

\subsection{Expansion of the density operator in driving speed} \label{subsec:expansion}

With finite speed driving, the equation of motion for the total system is
\begin{eqnarray} \label{eq:EOMfirst}
\frac {d} {dt} \hat  \rho (\bold R, t) =  \frac {\partial } { \partial t} \hat  \rho  + \sum_\nu \dot{R}^\nu  \partial_\nu \hat \rho = - \frac{i}{\hbar} [ \hat H (\bold R), \hat  \rho] 
\end{eqnarray}
or 
\begin{eqnarray} \label{eq:EOM1}
\frac {\partial } { \partial t} \hat  \rho (\bold R, t) = - \sum_\nu \dot{R}^\nu  \partial_\nu \hat \rho  - \hat{\hat{\mathcal{L}}} \hat \rho  \\
 - \hat{\hat{\mathcal{L}}} \hat  \rho \equiv - \frac{i}{\hbar} [ \hat  H, \hat  \rho]  \label{eq:EOM2}
\end{eqnarray}
where we have used Eq. \eqref{eq:first} to express the total time derivative of $\hat \rho$ (i.e. $d \hat \rho/dt $) as a combination of the the explicit contribution $\partial \hat \rho/ \partial t $ (which remains when the parameters $\bold R $ are constants) and the term(s) associated with time evolution of the parameters $\bold R$. The solution of Eq. \eqref{eq:EOM1} is then written as a power series in the driving speed:
\begin{eqnarray} \label{eq:rho123}
\hat \rho =\hat  \rho^{(0)} + \hat  \rho^{(1)}  + \hat  \rho^{(2)} + \cdots 
\end{eqnarray}
where $\hat  \rho^{(n)}$ represent contribution of order n in $\dot {\bold R}$. Substituting Eq. \eqref{eq:rho123} into Eq. \eqref{eq:EOM1} and matching orders of   $\dot {\bold R}$ from both sides, we get a series of equations, 
\begin{eqnarray} \label{eq:EOM0}
\frac {\partial } { \partial t}  \hat  \rho^{(0)}  &=& - \frac{i}{\hbar} [ \hat H, \hat  \rho^{(0)} ], \\
\frac {\partial } { \partial t} \hat  \rho^{(1)}  &=& - \frac{i}{\hbar} [\hat  H,  \hat  \rho^{(1)} ]  - \sum_\nu \dot{R}^\nu  \partial_\nu \hat  \rho^{(0)}, \label{eq:EOMrho1} \\
\frac {\partial } { \partial t} \hat  \rho^{(n)}  &=& - \frac{i}{\hbar} [ \hat  H, \hat  \rho^{(n)} ]  - \sum_\nu \dot{R}^\nu  \partial_\nu \hat  \rho^{(n-1)}, \: n>1.
\end{eqnarray}

Under the assumption that the dynamics represented by Eq. \eqref{eq:EOM0} is much faster than the time evolution of the parameters $\bold R$, the equilibrium solution of Eq. \eqref{eq:EOM0} can be used as a ``boundary condition" defining the inhomogeneous term in Eq. \eqref{eq:EOMrho1} \cite{PhysRevB.97.064303}, so that we can then proceed to solve for   $\hat  \rho^{(1)} $ (and $\hat  \rho^{(n)}, \: n>1 $): 
\begin{eqnarray} \label{eq:rhon1r}
 \hat \rho^{(1)} (\bold R, t) &=&   - \sum_\nu    \int_0^t  e^{-i \hat  H (t-t')/\hbar} \dot{R}^\nu \partial_\nu  \hat  \rho^{(0)} e^{i \hat H (t-t')/\hbar}  dt' \\
 \hat \rho^{(n)} (\bold R, t) &=&   - \sum_\nu    \int_0^t  e^{-i \hat  H (t-t')/\hbar} \dot{R}^\nu \partial_\nu \hat  \rho^{(n-1)} e^{i \hat H (t-t')/\hbar}  dt'; \: n>1 \label{eq:rhonad}
\end{eqnarray}
Note that as it stands, Eq. \eqref{eq:EOM0} does not have a unique steady state solution, as any function of $\hat H$ provides such a solution. Choosing the equilibrium solution $\hat \rho^{(0)} (\bold R) = e^{-\beta (\hat H (\bold R) -\mu \hat N )}/\Omega$  to generate the higher order terms in Eqs. \eqref{eq:rhon1r} and \eqref{eq:rhonad} is again based on the recognition that the actual time evolution is given by Eq. \eqref{eq:Xrho} that includes (small) coupling to an external equilibrium environment. This coupling brings the system to equilibrium for any given constant $\bold R$, however the coupling is assumed small enough so as not to affect the response of the system to the driving at any finite time. 

Note that, $\hat  \rho^{(0)}(\bold R) $ depends only on  $\bold R$ and does not depend on $t$ explicitly, whereas $\hat  \rho^{(n)} (\bold R, t)$ ($n \ge 1$) depends on $t$ explicitly. Note also that 
\begin{eqnarray}
{\rm Tr} \hat \rho = {\rm Tr} \hat \rho^{(0)} = 1, 
\end{eqnarray}
hence, 
\begin{eqnarray}
{\rm Tr} \hat \rho^{(n)} =0, \: n \geq 1. 
\end{eqnarray}
In what follows we makes another simplification, made possible by the nature of our problem. While the composite system under discussion is macroscopic (comprising the microscopic system of interest and its macroscopic environment), the changes represented by $\bold R$ are local, taking place within the microscopic subsystem or at its boundary (that is, in its coupling to the rest of the full system). The evolution  $e^{-i \hat  H (t-t')/\hbar} \dot{R}^\nu \partial_\nu  \hat  \rho^{(0)} e^{i \hat H (t-t')/\hbar} $ under the full system Hamiltonian takes the deviation of  $\hat  \rho^{(0)}$  from equilibrium, caused by a change in $\bold R$, back to zero. Assuming that this relaxation is fast relative to the driving speed, we can make the Markovian approximation
\begin{eqnarray} \label{eq:rho1mark}
 \hat  \rho^{(1)} (\bold R) &\approx& - \sum_\nu \dot{R}^\nu  \int_0^\infty  e^{-i \hat H t'/\hbar} \partial_\nu \hat  \rho^{(0)} e^{i \hat  H t'/\hbar}  dt' \equiv  - \sum_\nu \dot{R}^\nu \hat{\hat{\mathcal{L}}}^{-1}  \partial_\nu \hat  \rho^{(0)}
\end{eqnarray}
We have denoted $\hat{\hat{\mathcal{L}}}^{-1} (\cdot)=\int_0^\infty  e^{-i \hat H t'/\hbar} (\cdot) e^{i \hat  H t'/\hbar}  dt' $. \cite{PhysRevLett.119.046001} Note that, with a constant $\dot{\bold R}$, $\hat  \rho^{(1)}$ in this Markovian limit only depends on $\bold R$ (not on $t$ explicitly). Consequently, the {\em n}th order correction is, under the same Markovian assumption
\begin{eqnarray}
 \hat  \rho^{(n)} (\bold R) \approx  - \sum_\nu \dot{R}^\nu \hat{\hat{\mathcal{L}}}^{-1}  \partial_\nu \hat  \rho^{(n-1)} =  (- \sum_\nu \dot{R}^\nu \hat{\hat{\mathcal{L}}}^{-1}  \partial_\nu)^n \hat  \rho^{(0)}, \: n>1
\end{eqnarray}
Henceforward, we will refer to results associated with $\hat  \rho^{(1)}$ ``nonadiabatic".

\subsection{Nonadiabatic thermodynamics: second order in driving speed}

With the results above, let us now calculate the rate of change of the thermodynamic quantities up to the second order in driving speed by replacing $\hat  \rho^{(0)}$ in Eqs. \eqref{eq:E1}-\eqref{eq:N1} with $\hat \rho^{(1)} $ (Eq. \eqref{eq:rho1mark}) \cite{ochoa2018quantum}, 
\begin{eqnarray} \label{eq:E2}
\dot{E}^{(2)} = - \sum_{\alpha\nu} \dot{R^{\alpha}}\dot{R^{\nu}} {\rm Tr}( \partial_\alpha \hat H  \hat{\hat{\mathcal{L}}}^{-1}  \partial_\nu \hat \rho^{(0)}  ) -  \sum_{\alpha\nu} \dot{R^{\alpha}} \dot{R^{\nu}} {\rm Tr}( \hat  H \partial_\alpha  ( \hat{\hat{\mathcal{L}}}^{-1}  \partial_\nu \hat  \rho^{(0)} )), 
\end{eqnarray}
\begin{eqnarray} \label{eq:W2}
\dot{W}^{(2)} =  -\sum_{\alpha\nu} \dot{R^{\alpha}}\dot{R^{\nu}} {\rm Tr}( \partial_\alpha \hat  H  \hat{\hat{\mathcal{L}}}^{-1}  \partial_\nu \hat  \rho^{(0)}  ), 
\end{eqnarray}
\begin{eqnarray} \label{eq:Q2}
\dot{Q}^{(2)} =  - \sum_{\alpha\nu} \dot{R^{\alpha}} \dot{R^{\nu}} {\rm Tr}( \hat  H \partial_\alpha  ( \hat{\hat{\mathcal{L}}}^{-1}  \partial_\nu \hat  \rho^{(0)} )) +  \mu \sum_{\alpha\nu} \dot{R^{\alpha}} \dot{R^{\nu}} {\rm Tr}( \hat  N  \partial_\alpha   ( \hat{\hat{\mathcal{L}}}^{-1}  \partial_\nu \hat  \rho^{(0)} )), 
\end{eqnarray}
and 
\begin{eqnarray} \label{eq:N2}
\mu \dot{N}^{(2)}  = - \mu \sum_{\alpha\nu} \dot{R^{\alpha}} \dot{R^{\nu}} {\rm Tr}( \hat  N  \partial_\alpha  ( \hat{\hat{\mathcal{L}}}^{-1}  \partial_\nu \hat  \rho^{(0)} )) . 
\end{eqnarray}
Note that the second order work rate, Eq. \eqref{eq:W2}, is related to dissipation,
\begin{eqnarray} \label{eq:W2ab}
\dot{W}^{(2)} =  -\sum_{\alpha\nu} \dot{R^{\alpha}}\dot{R^{\nu}} {\rm Tr}( \partial_\alpha \hat  H  \hat{\hat{\mathcal{L}}}^{-1}  \partial_\nu \hat  \rho^{(0)}  ) = \sum_{\alpha \nu} \dot{R^{\alpha}} \gamma_{\alpha \nu}  \dot{R^{\nu}} ,
\end{eqnarray}
where the friction tensor $\gamma_{\alpha \nu }$ is defined by \cite{PhysRevLett.119.046001,PhysRevB.96.104305,PhysRevB.97.064303,dou2018perspective}
\begin{eqnarray} \label{eq:friction}
\gamma_{\alpha \nu } = {\rm Tr}( \partial_\alpha \hat H  \hat{\hat{\mathcal{L}}}^{-1}  \partial_\nu \hat  \rho^{(0)} ).
\end{eqnarray}
Generalizing the result of Eq. \eqref{eq:firstLaw1}, we now find that the first law of thermodynamics is obeyed to the second order in $\dot{\bold R}$  (i.e. in the nonadiabatic limit):
\begin{eqnarray} \label{eq:FirstLaw2}
\dot{E}^{(2)} = \dot{W}^{(2)} + \dot{Q}^{(2)}+ \mu \dot{N}^{(2)}.
\end{eqnarray}

As already alluded to at the end of Sec. \ref{sec:static} and following Eq. \eqref{eq:rhonad}, the heat and particle current are characteristics of the openness of the full system to the ``superbath" that determines the temperature and chemical potential of the equilibrium system. As discussed above, this information enters through the imposed form of  $\hat  \rho^{(0)} $ and does not explicitly depend on the coupling to this ``superbath". These results for rates of change of  the thermodynamic functions is mathematically consistent, but their physical interpretation should be assessed carefully as further discussed below.

\subsection{ Entropy production}
 Next consider entropy production. At equilibrium, the von Neumann entropy, Eq. \eqref{eq:S}, is the proper extension of the Gibbs entropy to quantum statistical thermodynamics. Here, as in Refs. \cite{PhysRevB.93.115318} and \cite{bruch2018landauer}  we explore the use of the same concept to slowly driven non-equilibrium systems by simply replacing the static density operator $\hat  \rho^{(0)} $  by the full density operator  $\hat  \rho$ (Eq. \eqref{eq:rho123})
\begin{eqnarray} \label{eq:fulS}
S =  - k_B  {\rm Tr}(\hat \rho \ln \hat \rho ) 
\end{eqnarray}
 The first order correction to $S^{(0)}$ is (see Appendix \ref{app:SSS}) 
\begin{eqnarray} \label{eq:Sad}
S^{(1)} =  - k_B  {\rm Tr}(\hat  \rho^{(1)} \ln \hat  \rho^{(0)} )   
\end{eqnarray}
or
\begin{eqnarray} \label{eq:Sad2}
S^{(1)} &=&  \frac1T  {\rm Tr}(\hat \rho^{(1)} \hat H ) - \mu \frac1T  {\rm Tr}(\hat \rho^{(1)} \hat N ) \nonumber \\
&=& - \frac1T  \sum_\nu  \dot{R^{\nu}}   {\rm Tr}(\hat H  \hat{\hat{\mathcal{L}}}^{-1}  \partial_\nu \hat \rho^{(0)} )   +  \mu \frac1T  \sum_\nu  \dot{R^{\nu}}   {\rm Tr}(\hat N  \hat{\hat{\mathcal{L}}}^{-1}  \partial_\nu \hat \rho^{(0)} )  
\end{eqnarray}
Finally, we take the derivative of $S^{(1)}$  with respect to time in order to calculate the rate of change of the entropy to second order in driving speed, 
\begin{eqnarray} \label{eq:Sad2}
\dot{S}^{(2)} =&& \sum_\alpha \dot{R^{\alpha}} \partial_\alpha S^{(1)} =- \frac1T  \sum_{\alpha \nu}  \dot{R^{\alpha}} \dot{R^{\nu}}   {\rm Tr}( \partial_\alpha \hat H  \hat{\hat{\mathcal{L}}}^{-1}  \partial_\nu \hat \rho^{(0)} ) \nonumber \\ &&- \frac1T  \sum_{\alpha \nu}  \dot{R^{\alpha}} \dot{R^{\nu}}   {\rm Tr}(\hat H   \partial_\alpha  (\hat{\hat{\mathcal{L}}}^{-1}  \partial_\nu \hat \rho^{(0)} ) )  
 + \mu \frac1T  \sum_{\alpha \nu}  \dot{R^{\alpha}} \dot{R^{\nu}}   {\rm Tr}( \hat N   \partial_\alpha  (\hat{\hat{\mathcal{L}}}^{-1}  \partial_\nu \hat \rho^{(0)} )
\end{eqnarray}
With Eqs. \eqref{eq:W2}-\eqref{eq:Q2}, we find
\begin{eqnarray} \label{eq:SQW}
\dot{S}^{(2)} =   \frac {\dot{Q}^{(2)}} T +  \frac {\dot{W}^{(2)}} T 
\end{eqnarray}
For one electronic (or bosonic) bath, $\gamma_{\alpha \nu }$, Eq. \eqref{eq:friction}, is positive definite \cite{PhysRevLett.119.046001,dou2018perspective}, so that the second law of thermodynamics is satisfied, 
\begin{eqnarray}\label{eq:frictionW}
 \dot{S}^{(2)} -  \frac {\dot{Q}^{(2)}} T = \frac {\dot{W}^{(2)}} T  =  \frac1T \sum_{\alpha \nu}  \dot{R^{\alpha}} \gamma_{\alpha \nu }  \dot{R^{\nu}}   \geq 0 
\end{eqnarray}

The relationship (Eq. \eqref{eq:SQW}), which has been reported previously in the literature \cite{PhysRevLett.114.080602, PhysRevB.93.115318, ochoa2018quantum}, indicates that the extension of Eq. \eqref{eq:fulS} for the entropy to driven non-equilibrium systems is consistent with our understanding of the entropy concept, in particular the association of entropy production with the (positive definite) energy dissipation, at least up to the second order in the driving speed. Several other points should be noted:

(a) Using Eq. \eqref{eq:FirstLaw2}, Eq. \eqref{eq:SQW} may be rewritten in the form 
\begin{eqnarray} \label{eq:SecondLawAgain}
\dot{E}^{(2)} - \mu \dot{N}^{(2)} = T \dot{S}^{(2)} 
\end{eqnarray}
The corresponding first order relation (Eq. \eqref{eq:firstLaw1} with $\dot Q^{(1)} = T \dot S^{(1)}$  ) is
\begin{eqnarray} \label{eq:firstLawAgain22}
\dot{E}^{(1)} - \mu \dot{N}^{(1)} = T \dot{S}^{(1)} + \dot{W}^{(1)}  \label{eq:firstLawAgain1} 
\end{eqnarray}
Indeed, Eqs.  \eqref{eq:SecondLawAgain} and \eqref{eq:firstLawAgain22} can be derived starting from the total averaged energy and particle number written in terms of the full density operator, 
\begin{eqnarray} \label{eq:fullE}
E - \mu N = {\rm Tr}( \hat \rho (\hat H-\mu \hat N)  ) = -\frac1\beta {\rm Tr}(\hat \rho \ln e^{-\beta (\hat H-\mu \hat N) } ) \nonumber \\
= -\frac1\beta {\rm Tr}( \hat \rho \ln \hat \rho^{(0)}  )  - \frac1\beta \ln \Omega 
\end{eqnarray}
With Eq. \eqref{eq:rho123}, the above equation gives
\begin{eqnarray} \label{eq:E0mN0}
E^{(0)} - \mu N^{(0)} &=& {\rm Tr}( \hat \rho^{(0)} (\hat H-\mu \hat N)  ) =-\frac1\beta {\rm Tr}(\hat \rho^{(0)} \ln \hat  \rho^{(0)}  )  - \frac1\beta \ln \Omega     \\
E^{(n)} - \mu N^{(n)} &=&  {\rm Tr}( \hat \rho^{(n)} (\hat H-\mu \hat N)  )  = -\frac1\beta {\rm Tr}(\hat \rho^{(n)} \ln \hat  \rho^{(0)}  )  \: n \geq 1 \label{eq:EnmNn}
\end{eqnarray}
From Eq. \eqref{eq:S}, Eq. \eqref{eq:E0mN0} is just $E^{(0)} - \mu N^{(0)} =T S^{(0)} - \frac1\beta \ln \Omega$ , while using Eq. \eqref{eq:Sad} and ${\rm Tr} \hat \rho^{(1)} =0$, the $n=1$ equation of Eq. \eqref{eq:EnmNn} is $E^{(1)} - \mu N^{(1)} =T S^{(1)} $. The time derivatives of these two equations yields Eq. \eqref{eq:firstLawAgain22} (since $\dot W^{(1)}=dF^{(0)}/dt = - \beta^{-1} d/dt \ln \Omega$) and Eq. \eqref{eq:SecondLawAgain}, respectively.

(b) Beyond second order, however, while  $E^{(n)} - \mu N^{(n)}   = -\frac1\beta {\rm Tr}(\hat \rho^{(n)} \ln \hat  \rho^{(0)}  ) $ we cannot relate this expression to the corresponding order of the entropy expansion. Namely, while such an expansion can be formally obtained from Eqs. \eqref{eq:fulS} and  \eqref{eq:rho123}, we find that 
\begin{eqnarray}
S^{(n)} \ne - k_B {\rm Tr}(\hat \rho^{(n)} \ln \hat \rho^{(0)} ), \:  n\geq 2 
\end{eqnarray}
 so that  $ E^{(n)} - \mu  N^{(n)}  \ne T S^{(n)}, n\geq 2$. Consequently
\begin{eqnarray}
\dot E^{(n+1)} - \mu \dot N^{(n+1)}&& \ne T \dot S^{(n+1)}, \: n\geq 2 \label{eq:fristLawAgain3}
\end{eqnarray}
so that this procedure, which relies on the definition Eq. \eqref{eq:fulS} appears to fail beyond second order.

(c) In the quasi-static limit  $\dot S^{(1)}=\dot Q^{(1)}/T$, which tells us the the change of entropy in the full system is essentially given by the heat flux into the system. The departure from this relationship at the next (second) order expresses the fact that in addition to heat flux, there is an additional source of entropy -- the dissipated work $\dot W^{(2)}$. The latter is identified as the entropy production,  $\dot S^{(2)} -\dot Q^{(2)} /T=\dot W^{(2)} /T>0$. Note that in Ref. \cite{bruch2018landauer} the same physics was expressed from the outside perspective: The outwards entropy flux was shown to be smaller than the outward heat flux divided by $T$ by the amount $\dot W^{(2)}/T$,  which expresses the increase in entropy remaining in the system due to the dissipated work.

(d) As another way to look at the thermodynamics of the driven system, consider the equilibrium states 1 and 2 that correspond to two sets of system parameters,  $\bold R_1$ and $\bold R_2$, respectively. Starting from state 1 consider a protocol $\bold R(t)$  that eventually takes the system to state 2. Since both states 1 and 2 are well defined equilibrium states, the change in any state function $\mathcal{F}$ is independent of the protocol and can be calculated from the integral over the quasi-static process,  $\Delta \mathcal{F} = \mathcal{F}(2) - \mathcal{F}(1) = \int_1^2 dt \dot{\mathcal{F}} (t) $. Consequently, for any driving protocol, any state function must satisfy $\delta \mathcal{F}= 0$, where the excess function $\delta \mathcal{F} $  is defined by  $\delta \mathcal{F} =\mathcal{F} (t)- \mathcal{F}^{(1)}(t)$ The work obviously depends on the $\bold R(t)$ protocol and is given by the first term on the RHS of Eq. \eqref{eq:dE_dtful},  $W=\int_1^2 dt {\rm Tr}(\hat \rho \sum_\alpha \dot R^\alpha \partial_\alpha \hat H )$. The excess work, $\delta W=W - W^{(1)}$  is the dissipated work associated with the irreversible driving. In particular, to second order in the driving speed, dissipated work is given by $W^{(2)}=\int_1^2 dt \sum_{\alpha\nu} \dot R^\alpha \gamma_{\alpha\nu} \dot R^\nu$. The first law implies that this excess work equals the excess heat that is given to the ``superbath" during the process,  $\delta W= - \delta Q = - (Q-Q^{(1)})$, and consquently 
\begin{eqnarray} \label{eq:S2equi}
\delta S = \int_1^2 dt \dot S^{(2)} =0, 
\end{eqnarray}
in agreement with Eq. \eqref{eq:SQW}.  

{  The apparent contrast between Eq. \eqref{eq:S2equi} and  Eq. \eqref{eq:SQW} needs be clarified: Eq. \eqref{eq:S2equi} states the obvious -- the system entropy change between two equilibrium states is fully accounted for by the corresponding quasi-static process, irrespective of driving protocol; whereas Eq. \eqref{eq:SQW} quantifies the instantaneous rate of system entropy change due to dissipative processes along a trajectory on which the system is driven at finite-speed.} First, note that $\int_1^2 dt \dot S^{(2)} \ne 0$   if the integral is done between any two points along the finite-speed trajectory. It vanishes only between equilibrium points, i.e. when the driving came to rest and enough time has passed to allow the system to equilibrate. Secondly, in the latter case, when states 1 and 2 are equilibrium states, the fact that $\int_1^2 dt \dot S^{(2)} =0$ for any driving protocol used to induce the $1\rightarrow 2$ process only means that any entropy produced in the process is associated with the heat transferred to the external ``superbath". The excess entropy produced when this protocol induces irreversible dynamics (such as with finite speed driving but including the relaxation that takes place after the driving stops  until the system comes to complete equilibrium) can be identified as $T^{-1} \delta Q = T^{-1} \int_1^2 dt \dot Q^{(2)} $, provided that  $\dot Q^{(2)}$ describes also the heat transferred to/from the superbath during this relaxation segment.

In light of these remarks, the physical contents of Eq. \eqref{eq:SQW} can be understood as follows. In the expression $\dot S^{(2)}  =( \dot W^{(2)} + \dot Q^{(2)} )/T$, $\dot W^{(2)} >0 $ is the excess work done because of the finite speed driving. At the end, all this excess work will exit as heat to the superbath of temperature $T$, implying entropy change in the universe of  $T^{-1} \int_1^2 dt \dot W^{(2)} $. However, at any point in time  $-\dot Q^{(2)} $  is the rate of heat escaping the system into the superbath (note that our choice of sign is that positive $Q$ describes heat entering the system) and the difference  $\dot S^{(2)}  =( \dot W^{(2)} -  (-\dot Q^{(2)}) )/T$ describes entropy increase in the system. The total rate of entropy production  $\dot S^{(2)} +  (-\dot Q^{(2)} /T) $ -- the sum of rates of excess entropy generated in the system (calculated has the second order contribution, $\dot S^{(2)}$,  to $(d/dt) {\rm Tr}(\hat \rho \ln \hat \rho) $  (Eqs. (38)-(40)), and the rate of excess entropy produced in the ``superbath". In a change between equilibrium states no excess entropy is produced in the system, $\int_1^2 dt \dot S^{(2)} =0 $ , and all the excess work is dissipated as heat into the external ``superbath",  $\int_1^2 dt \dot W^{(2)} = - \int_1^2 dt \dot Q^{(2)} $.

	Our discussion so far has focused on the thermodynamics of the full system. Next we consider the thermodynamics of the interesting subsystem on which the driving is done: We assume that the parameters  $\bold R$ characterize this subsystem and/or its interaction with its environment.   

\section{system-bath separation} \label{sec:sub-system}
 
In the above treatment, the basis for the Markovian assumption, Eq. \eqref{eq:rho1mark}, was the local nature of the driving. Here we further explore this local nature by separating the full system considered above into a sub-system D (henceforth referred to as ``dot") and a bath B, with the Hamiltonian written as 
\begin{eqnarray} \label{eq:generalH}
\hat H &=& \hat H_{D} + \hat H_B + \hat H_{I} , 
\end{eqnarray}
where $\hat H_{I}$  is the coupling between the sub-system and bath. While the analysis above has focused on the effect of driving on the thermodynamics of the full D+B system, our aim now is to address, as usually done, the thermodynamic properties of the subsystem of interest -- the dot D. With this in mind we assume that the driving takes place within this subsystem, that is, $\hat H_D=\hat H_D(\bold R)$  while  $\hat H_B$ and $\hat H_I$   are constant. The dot and its driving dynamics can otherwise be general, with arbitrary number of levels and driving parameters.

\begin{figure}[htbp] %  figure placement: here, top, bottom, or page
   \centering
   \includegraphics[width=3in]{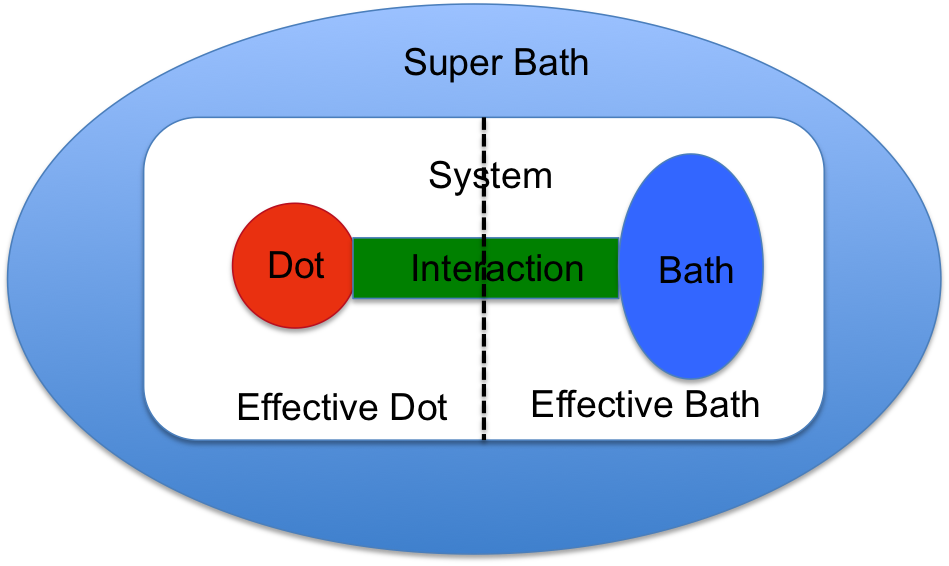} 
   \caption{Our system consists of a dot, a bath and interaction between the dot and the bath. With the proper splitting of the interaction in Eqs. \eqref{eq:fulHsplit1}-\eqref{eq:fulHsplit3}, the system is divided into a effective dot and a effective bath. A super bath may or may not be present around the total system.}
   \label{fig:system}
\end{figure}

In the weak sub-system--bath coupling regime, the distinction between sub-system and bath is based on two attributes: First, the `sub-system' is the focus of our interest (in the present case because it is the subject of the external driving and of any subsequent measurement) and second, it is assumed the sub-system--bath coupling is much weaker than the interactions that bring the bath to thermal equilibrium. Under these assumptions the full density operator is written as a direct product of the sub-system density operator and the density operator of the equilibrium bath:  $\hat \rho= \hat \rho_D \otimes \hat \rho_B^{eq}$ and the evolution of sub-system properties are obtained by evaluating  $\hat \rho_D$. In the strong coupling regime, however, sub-system and bath become entangled and such a decomposition of   $\hat \rho$ does not hold. Still, even in this case it would possible to consider separately the thermodynamic properties of the two subsystems provided that an unambiguous way exists for splitting the contribution of the interaction operator $\hat H_I$ between them. In general no such procedure exists, however the expectation values of {\em single particle} operators can be separated between the two subsystems based on the following consideration: The expectations values of such operators can be written as traces over single particle states involving the single particle density matrix. For example, if $\hat A$ is such an operator, $\hat A=\sum_{ij} A_{ij} \hat c^+_{i} \hat c_j$ (where $\hat c$ and $\hat c^+$ are single particle annihilation and creation operators), the expectation value of $\hat A$ can be written in the basis of single particle states of the free D and B systems,  respectively, in the form 
\begin{eqnarray} \label{eq:A_DB}
\langle \hat A \rangle = {\rm Tr} (\hat \rho \hat A) = \frac12 {\rm Tr} (\hat \rho \hat A + \hat A  \hat \rho) = \sum_{ij} \frac12 ( A_{ij} \sigma_{ji} + \sigma_{ij} A_{ji} ) \nonumber \\ 
= \sum_{i\in D, j} \frac12 ( A_{ij} \sigma_{ji} + \sigma_{ij} A_{ji} ) + \sum_{i\in B, j} \frac12 ( A_{ij} \sigma_{ji} + \sigma_{ij} A_{ji} )  \equiv \langle \hat A \rangle_D +\langle \hat A \rangle_B
\end{eqnarray}
{  The symmetric forms of $\langle \hat A \rangle_D$ and $\langle \hat A \rangle_B$ are needed to guarantee that these expectations are real-valued.}  $\sigma$  here is the single particle density matrix, $\sigma_{ij}={\rm Tr}(\hat \rho \hat c^+_j \hat c_i )$. This suggests a natural separation of averaged single particle observables into parts associated with the individual subsystems. When  $\hat A=\hat N= \hat N_D+ \hat N_B$ is the number operator, this expresses the trivial separability of the total particle number into a sum of particle numbers in the two subsystems. When $\hat A$  is a Hamiltonian of a non-interacting fermion or boson model, that is Eq. \eqref{eq:generalH} with $\hat H_D = \sum_d \epsilon_d \hat c^+_d \hat c_d $, $\hat H_B = \sum_b \epsilon_b \hat c^+_b \hat c_b $, $\hat H_I = \sum_{db} V_{bd} (\hat c^+_d \hat c_b + \hat c^+_b \hat c_d) $, we have $\langle \hat H \rangle = {\rm Tr} (\hat \rho \hat H) = \langle \hat H \rangle_D + \langle \hat H \rangle_B $ with (similar to Eq. \eqref{eq:A_DB}), 
\begin{eqnarray} \label{eq:E_Dnon1}
\langle \hat H \rangle_D  =  \sum_d \sigma_d \rho_{dd} +  \frac12 \sum_{bd} (V_{db} \sigma_{db}+ \sigma_{db} V_{db}  ) = \langle \hat H_D \rangle  + \frac12 \langle \hat H_I \rangle 
\end{eqnarray}
and
\begin{eqnarray}\label{eq:E_Dnon2}
\langle \hat H \rangle_B =  \sum_b \sigma_b \rho_{bb} +  \frac12 \sum_{bd} (V_{db} \sigma_{db}+ \sigma_{db} V_{db}  ) = \langle \hat H_B \rangle  + \frac12 \langle \hat H_I \rangle \nonumber \\
\end{eqnarray}
where $\langle \hat H_I \rangle= {\rm Tr}(\hat \rho \hat H_I )$.
Thus, the assumption that the interaction energy between the the two subsystem is evenly split between them, assumed in several recent papers\cite{PhysRevB.93.115318,ochoa2018quantum}, naturally holds  in models of non-interacting bosons or fermions. \cite{thermodynamics:Hs} Fig. \ref{fig:system} provides a schematic view of this splitting. It should be emphasized, however, that this result holds only for a restricted set of models and for expectation values of operators with bilinear system-bath coupling. In general, it is not true that $\hat H_D + \frac12 \hat H_I $  and $\hat H_B + \frac12 \hat H_I $  are effective Hamiltonians for the two subsystems; these effective Hamiltonians can be used to calculate first moments of the energy distribution, but will in general fail capturing higher moments. \cite{PhysRevB.94.035420}

Assuming that this even splitting of  $\langle \hat H_I \rangle$ between the D and B system holds, we can now consider the thermodynamic properties of each subsystem. In particular we focus on the dot D. Note that, when considering the full D+B system we must resort to the (at least conceptual) existence of a ``superbath" that maintains the equilibrium properties of the (otherwise closed) system, but at the same time, the dynamics at the D-B interface should not depend on the existence of such a superbath. Thus, when we focus now on the subsystem D, we will proceed by considering D+B as a closed system, keeping in mind that B is infinitely large. For this closed system we can write
\begin{eqnarray} \label{eq:fulHsplit1}
\hat H = \hat H_D^{eff} + \hat H_B^{eff} \\
\hat H_D^{eff} = \hat H_D+ \frac12\hat H_I \label{eq:fulHsplit2}\\
\hat H_B^{eff} = \hat H_B + \frac12\hat H_I\label{eq:fulHsplit3}
\end{eqnarray}
which is assumed to hold as long as we limit ourselves to the calculation of first moments of these operators. Furthermore, obviously  $\hat N= \hat N_D + \hat N_B $.

{\bf First law}.  We can now consider the energy change in the full system
\begin{eqnarray} \label{eq:dEdteff}
\frac{dE}{dt} = {\rm Tr} (\frac{d \hat H }{dt} \hat \rho) + {\rm Tr} (\hat H \frac{d \hat \rho }{dt} )= {\rm Tr}  (\frac{d \hat H_D }{dt} \rho) + {\rm Tr}  ( (\hat H_D^{eff}+ \hat H_B^{eff}) \frac{d \hat \rho }{dt} )
\end{eqnarray}
and its trivial (since $d\hat N/dt =0$ for this closed system) extension
\begin{eqnarray} \label{eq:dE-muN}
\frac{d (E - \mu N )}{dt} = {\rm Tr}  (\frac{d \hat H }{dt} \hat \rho) + {\rm Tr} ( (\hat H-\mu \hat N) \frac{d \hat \rho }{dt} ) \nonumber \\
= {\rm Tr} (\frac{d \hat H_D }{dt} \hat \rho) + {\rm Tr}  ( (\hat H_D^{eff}+ \hat H_B^{eff} - \mu (\hat N_D+ \hat N_B) ) \frac{d \hat \rho }{dt} )
\end{eqnarray}
The last terms on the RHS of Eq. \eqref{eq:dEdteff} and Eq. \eqref{eq:dE-muN} vanish for the full closed system (see Eq. \eqref{eq:dQ/dtful} and subsequent text). However, these expressions can be separated into their dot and bath parts. In particular
\begin{eqnarray} \label{eq:dotE_Dful}
\dot E_D = \mu \dot N_D  +  \dot W_D + \dot Q_D
\end{eqnarray}
where 
\begin{eqnarray} \label{eq:W_Dful}
\dot W_D = {\rm Tr}(\frac{d\hat H_D^{eff} }{dt} \hat \rho ) = {\rm Tr}(\frac{d\hat H_D}{dt} \hat \rho) \\
\dot N_D = {\rm Tr} (\hat N_D \frac{d\hat \rho}{dt}  ) \\
E_D = {\rm Tr}(\hat H_D^{eff}  \hat \rho ) 
\end{eqnarray}
and  
\begin{eqnarray}\label{eq:Q_Dful}
\dot Q_D = {\rm Tr}( ( \hat H_D^{eff}-\mu \hat N_D  ) \frac{d \hat \rho}{dt} ) , 
\end{eqnarray}
thus establishing the first law of thermodynamics for the subsystem D. This result stems only from the the separability expressed by Eqs. \eqref{eq:fulHsplit1}-\eqref{eq:fulHsplit3}. For slow driving we can further apply the expansion (Eq. \eqref{eq:rho123}) of the full density operator $\hat \rho$  in powers of the driving speed  $\dot {\bold R}$,  $\hat \rho=\hat \rho^{(0)} + \hat \rho^{(1)} + \cdots$. and rewrite Eq. \eqref{eq:dotE_Dful} in the corresponding orders 
\begin{eqnarray}
\dot E_D^{(1)} = \mu \dot N_D^{(1)}  +  \dot W_D^{(1)} + \dot Q_D^{(1)} \\
\dot E_D^{(2)} = \mu \dot N_D^{(2)}  +  \dot W_D^{(2)} + \dot Q_D^{(2)} \label{eq:E_D2rd}
\end{eqnarray}
where the first and second order rates are obtained by replacing  $\hat \rho$ by $\hat \rho^{(0)}$  and $\hat \rho^{(1)}$   in Eqs. \eqref{eq:W_Dful}-\eqref{eq:Q_Dful}, respectively. Note that if only $\hat H_D$  is changing by the driving,  $\dot W_D^{(2)} = \dot W^{(2)} \ge 0$ (cf. Eq. \eqref{eq:frictionW}).

	While the work term in the above expressions is conceptually straightforward, the physical contents of the heat term is less obvious. Note that (since $\dot Q_D+\dot Q_B=0$, see Eq. \eqref{eq:dQ/dtful})
\begin{eqnarray} \label{eq:Q_DfulAgain}
\dot Q_D = Tr((\hat H_D^{eff}-\mu \hat N_D)  \frac{d\hat \rho}{dt} ) = - \dot Q_B = -Tr((\hat H_B^{eff}-\mu \hat N_B)  \frac{d\hat \rho}{dt} )
\end{eqnarray}
so these rates represent a heat current between dot and bath induced by the driving. It is also important to note that although these relationships are derived from a closed full system (D+B) picture, the appearance of  $\mu$ indicates that eventually particles of the full system are exchanged with a superbath of chemical potential $\mu$. Indeed, Eq. \eqref{eq:Q_DfulAgain} is a bookkeeping device: particles move between sub-systems D and B so that the total energy and number of particles are conserved. However, Eq. \eqref{eq:Q_DfulAgain} is a statement that particles exchanged with the bath will eventually, even if on a very different timescale, be exchanged with a superbath of chemical potential $\mu$. 

{\bf Entropy}. The entropy of the full closed D+B system, $S=-k_B {\rm Tr}(\hat \rho \ln \hat \rho)$, is conserved during its unitary evolution. This is easily shown explicitly, repeating the procedure outlined in Appendix \ref{app:SSS}: 
\begin{eqnarray}
\frac{d} {dt} S =  - k_B  {\rm Tr}(\frac{d \hat \rho}{ dt}   \ln \hat \rho ) -  k_B  {\rm Tr}( \hat \rho \hat \rho^{-1}  \frac{d \hat \rho}{ dt} ) =  \frac{i}{\hbar} k_B {\rm Tr}( [\hat H, \hat \rho]  \ln \hat \rho )   = 0 
\end{eqnarray}
This implies that, as in Eq. \eqref{eq:Q_DfulAgain}, if proper splitting of the entropy to its D and B parts can be formulated, changes in the subsystem entropies will reflect entropy flow between them(see also Ref. \cite{bruch2018landauer}). To define such partial entropies we use the already established splitting of the energy and number operators $\hat H$  and $\hat N$  and rewrite the time evolution $dS/dt$  in terms of these operators. To this end we use the expansion \eqref{eq:rho123} and the definition \eqref{eq:fulS} to write the corresponding expansion of $S$. 
\begin{eqnarray} \label{eq:dSdt1}
\frac{d} {dt} (S^{(0)} + S^{(1)} + \cdots) = - k_B \frac{d}{ dt} {\rm Tr}( (\hat  \rho^{(0)} +\hat  \rho^{(1)}  + \cdots ) \ln (\hat  \rho^{(0)} +\hat  \rho^{(1)}  + \cdots ) )  = 0 \nonumber \\
\end{eqnarray}
The first two terms were obtained above:  $S^{(0)} = - k_B  {\rm Tr}( \hat  \rho^{(0)} \ln \hat  \rho^{(0)} )$ (Eq.  \eqref{eq:S}) and $S^{(1)} = - k_B {\rm Tr}( \hat  \rho^{(1)} \ln \hat  \rho^{(0)} ) $  (Eq. \eqref{eq:Sad}). Using ${\rm Tr}(d\hat \rho/dt)=0$, the time derivative of the former is obtained as
\begin{eqnarray}
\dot S^{(1)} = \frac{d}{dt} S^{(0)} =  \frac{1}{T}  {\rm Tr}( (\hat  H -\mu \hat N) \frac{d }{dt} \hat  \rho^{(0)}) 
\end{eqnarray}
while that of $S^{(1)}$ is given by
\begin{eqnarray}
\dot S^{(2)} = \frac{d}{dt} S^{(1)} =  \frac{1}{T} \frac{d}{dt} {\rm Tr}( \hat  \rho^{(1)}  ( \hat H-\mu \hat N) ) 
\end{eqnarray}
At this point we have expressed the entropy  to first order in terms of $\hat H$  and $\hat N$, so that we can adopt the splitting of Eqs. \eqref{eq:fulHsplit1}-\eqref{eq:fulHsplit3}, and define the rate of entropy change in the sub-systems. In particular, 
\begin{eqnarray}
\dot S^{(1)}_D \equiv   \frac{1}{T}  {\rm Tr}( (\hat  H^{eff}_D -\mu \hat N_D) \frac{d }{dt} \hat  \rho^{(0)}) \\
\dot S^{(2)}_D \equiv  \frac{1}{T} \frac{d}{dt} {\rm Tr}( \hat  \rho^{(1)}  ( \hat H^{eff}_D -\mu \hat N_D) ) 
\end{eqnarray}
And using the definitions in Eq. \eqref{eq:dotE_Dful}-\eqref{eq:Q_Dful}, we arrive at 
\begin{eqnarray} \label{eq:S_D1st}
T\dot S^{(1)}_D = \dot Q^{(1)}_D = \dot E^{(1)}_D - \mu \dot N^{(1)}_D - \dot W^{(1)}_D  \\
T \dot S^{(2)}_D = \dot Q^{(2)}_D + \dot W^{(2)}_D  = \dot E^{(2)}_D - \mu \dot N^{(2)}_D  \label{eq:S_D2nd}
\end{eqnarray}
where again, the different orders of  $\dot Q$, $\dot W$ and $\dot N$  are obtained by substituting the corresponding orders of $\hat \rho$ in Eqs. \eqref{eq:W_Dful}-\eqref{eq:Q_Dful}.

Finally, as before, to the second order in the driving speed, the entropy change in subsystem D is seen to be associated with the friction work: When the driving affects only $\hat H_D$, the entropy production is given by $T\dot S_D^{(2)}-\dot Q_D^{(2)}=\dot W_D^{(2)} =\dot W^{(2)} \ge 0$. However, as noted before, this formalism cannot be extended in a simple way to higher orders in the driving speed.

\subsection{The resonant-level model}

Several recent papers have considered the driven resonant-level model as a simple test platform for quantum thermodynamics in strongly interacting situations. \cite{} Here we apply the formalism developed above to this model. The Hamiltonian is
 \begin{eqnarray} \label{eq:resonant}
\hat H = \epsilon_d (t) \hat d^+ \hat d + \sum_k V_k ( \hat d^+ \hat c_k  + \hat c_k^+ \hat d) + \sum_k \epsilon_k \hat c_k^+ \hat c_k  
\end{eqnarray}				
where the dot level (sub-system D with creation and annihilation operators now denoted  $\hat d^+$, $\hat d$) couples linearly to a manifold of electronic levels $k$ of the bath B through  $V_k$. The retarded self-energy of the dot level is defined to be
 \begin{eqnarray} \label{eq:SigmaE}
\Sigma(\epsilon) = \sum_k \frac{V_k^2}{ \epsilon-\epsilon_k + i\eta}  ,
\end{eqnarray}								
and the corresponding spectral function is
\begin{eqnarray}
A (\epsilon) = \frac{- 2 \mbox{Im} \Sigma (\epsilon) } { (\epsilon- \epsilon_d - \Re \Sigma (\epsilon) )^2+ (\mbox{Im} \Sigma(\epsilon) )^2}.
\end{eqnarray}
Applying the formalism of Sec. \ref{sec:sub-system} to this model, Eqs. \eqref{eq:W_Dful}-\eqref{eq:E_D2rd} and \eqref{eq:S_D1st}-\eqref{eq:S_D2nd} lead to explicit expressions for the different rates. In particular, the non-adiabatic correction to the work per unit time done to drive the system (frictional work) is obtained as (Appendix A)
\begin{eqnarray} \label{eq:W2A}
 \dot W^{(2)}_{D}  
= - \frac{\hbar \dot \epsilon_d^2}2  \int \frac{d\epsilon}{2\pi} A^2   \frac { \partial f(\epsilon) } { \partial \epsilon  } \end{eqnarray}
where $f(\epsilon)=(1+\exp(\beta (\epsilon-\mu))^{-1}$  is the Fermi distribution. It is obviously positive, satisfying the general result Eq. \eqref{eq:frictionW}. 
The corresponding entropy change is given by (Appendix B)
\begin{eqnarray} \label{eq:E2A}
T \dot S^{(2)}_{D} = \dot E^{(2)}_{D} -  \mu \dot N^{(2)}_{D} = -\frac{\hbar \dot \epsilon_d^2}2  \int \frac{d\epsilon}{2\pi} (\epsilon - \mu) \frac{\partial A^2 }{\partial \epsilon_d } \frac {\partial f(\epsilon)} {\partial \epsilon  }  
\end{eqnarray}
can be shown (Appendix B) to satisfy the relationship (compare Eq. \eqref{eq:SQW})
\begin{eqnarray}
T \dot S^{(2)}_{D} - \dot Q^{(2)}_{D}  = \dot W^{(2)}_{D}  
\end{eqnarray}
from which the heat flux may be obtained.
These results are obtained without invoking the wide-band approximation. If we further make this approximation the retarded self energy (Eq. \eqref{eq:SigmaE}) becomes pure imaginary and independent of  $\epsilon$, $\Sigma(\epsilon)=- \frac{i\Gamma}2$. In this limit the heat current can be simplified to give (Appendix B) 
\begin{eqnarray}
\dot Q^{(2)}_{D}  =   \dot E^{(2)}_{D} - \mu \dot N^{(2)}_{D} - \dot W^{(2)}_{D} = -\frac{\hbar \dot \epsilon_d^2}2   \int \frac{d\epsilon}{2\pi}   (\epsilon - \mu) A^2   \frac {\partial^2 f(\epsilon)} {\partial \epsilon^2  }   
\end{eqnarray}
The corresponding results of $\dot W^{(2)}_{D} $, $\dot Q^{(2)}_{D}$ and $\dot S^{(2)}_{D}$ are in agreement with the results in Ref. \cite{PhysRevB.93.115318}.

\subsection{The Anderson model}
 The general framework of Sec. \ref{sec:static}-\ref{sec:nonadiabatic} does not depend on the details of the system considered, and is applicable regardless of whether systems of free or interacting particles are considered. Such details are of course important for actual calculations of the thermodynamic functions. The simplest generalization of the resonant level model (Eq. \eqref{eq:resonant}) to include electron-electron interaction is the Anderson model  
\begin{eqnarray} \label{eq:andersonH}
\hat H = \epsilon_d (t) \sum_\sigma \hat d^+_\sigma \hat d_\sigma + U \hat d^+_\uparrow \hat d_\uparrow \hat d^+_\downarrow \hat d_\downarrow+ \sum_{k\sigma} V_k ( \hat d^+_\sigma \hat c_{k\sigma}  + \hat c_{k\sigma}^+ \hat d_\sigma) + \sum_{k\sigma} \epsilon_{k} \hat c_{k\sigma}^+ \hat c_{k\sigma}  
\end{eqnarray}
where now we include spin degrees of freedom,  $\sigma=\uparrow, \downarrow$, explicitly. 

As discussed above, once we go beyond non-interacting particle models, the splitting, Eq. \eqref{eq:E_Dnon1}-\eqref{eq:E_Dnon2} of the system bath interaction energy between system and bath does not hold rigorously. The calculation described below is based on the assumption that imposing such splitting, namely defining system and bath Hamiltonians by Eqs. \eqref{eq:fulHsplit1}-\eqref{eq:fulHsplit3} for the purpose of calculating average energies is still a reasonable approximation. Note that the calculation of the friction $\gamma$ and the corresponding excess work $\dot W_D^{(2)}$ (Ref. \cite{PhysRevLett.119.046001}) does not require this splitting assumption. It is however needed for evaluating (or rather assigning) the thermodynamic energies associated with the dot subsystem. 

In the calculation described below, we set  
\begin{eqnarray} \label{eq:gxunit}
\epsilon_d (t) = \epsilon_0+ \sqrt{2} g x(t)
\end{eqnarray}
where $x$  changes with time.  In order to calculate the thermodynamic quantities we need to diagonalize the Anderson Hamiltonian Eq. \eqref{eq:andersonH}, which can be done through numerical renormalization group (NRG) theory. For simplicity, we will apply the wide band approximation, such that $\Gamma=2\pi\sum_k V_k^2 \delta(\epsilon-\epsilon_k)$  is a constant. The details behind an NRG calculation can be found in Ref. \citenum{nrgreview} and in the supplemental material of Ref. \citen{PhysRevLett.119.046001}. We also present crucial steps in Appendix D.

Here we use this model to calculate  $E_{D}^{(1)}-\mu N_{D}^{(1)}$, where
\begin{eqnarray}
E_{D}^{(1)}-\mu N_{D}^{(1)} = {\rm Tr}( (\hat H^{eff}_{D}-\mu \hat N_{D})  \hat \rho^{(1)} ) 
\end{eqnarray}
$\dot E_D^{(2)}-\mu \dot N_D^{(2)}$ is then simply the derivative of $E_{D}^{(1)}-\mu N_{D}^{(1)}$ with respect to $t$ (or $x$, or $\epsilon_d$ since we have defined $\epsilon_d (t) = \epsilon_0+ \sqrt{2} g x(t)$).
Without loss of generality, we will set  $\mu=0$, such that it is sufficient to calculate $E_D^{(1)}={\rm Tr}( \hat H_D^{eff}  \hat \rho^{(1)} )$. 

\begin{figure}[htbp] %  figure placement: here, top, bottom, or page
   \centering
   \includegraphics[width=4in]{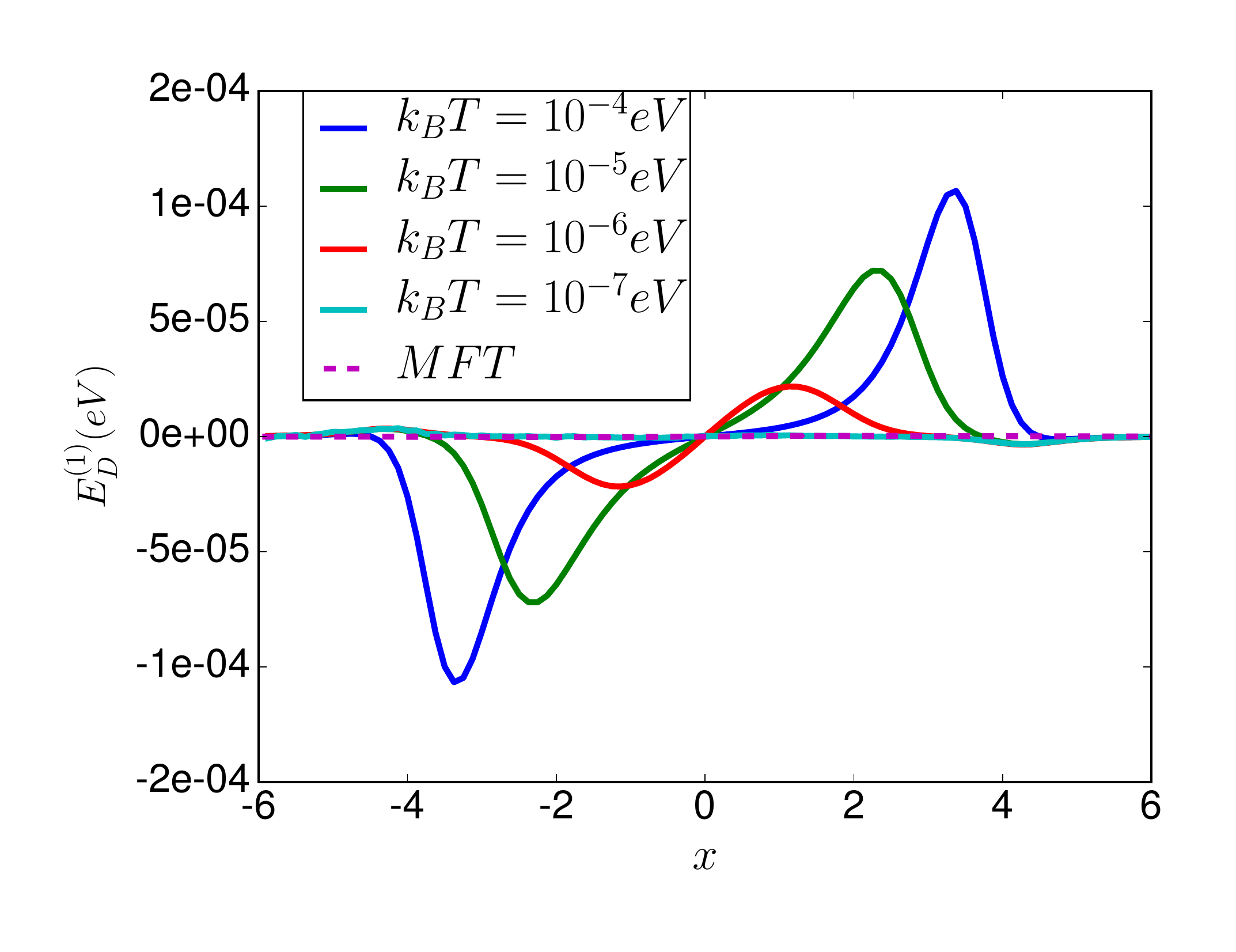} 
   \caption{With electron-electron interactions, $E_{D}^{(1)}$ from MFT (dashed line) is nearly zero. The MFT result is plotted at $k_BT=10^{-4}eV$. At lower temperature, MFT results are even smaller (not shown). By contrast,  NRG (solid lines) predicts notable peaks. Just as in Ref. \cite{PhysRevLett.119.046001}, these NRG peaks shift with temperature and are due to Kondo resonances, where the position of the peaks corresponds to the case where Kondo temperature is comparable with the actual temperature. At very low temperature, the Kondo peaks vanish. Parameters: $U=0.1eV$, $\Gamma=0.01eV$, $\epsilon_0=-0.05eV$, $g=0.0075eV$, $\mu=0$. We also set $\hbar \dot x=0.001eV$ ({  $x$ is unitless, see Eq. \eqref{eq:gxunit}}). }
   \label{fig:NRG}
\end{figure}

As shown in Fig. \ref{fig:NRG}, $E_{D}^{(1)}$ is nearly zero when electron-electron interactions are treated within a mean-filed theory (MFT, see Appendix D). That being said, when these interactions are treated within NRG, we see notable peaks. Furthermore,  these NRG peaks shift with temperature. Just as in Ref. \citen{PhysRevLett.119.046001}, these peaks arise due to Kondo resonances, and the peak positions reflect a match-up between the Kondo temperature and the actual temperature. At very low temperature, these Kondo peaks vanish. Understanding such Kondo signatures and their behaviors will require further analytical theory and investigation in the future.\cite{LangrethPRB1998,LangrethPRB1999} 

\section{Conclusion} \label{sec:con}
We have used a density operator formalism to investigate the quantum thermodynamics of a driven system. Such a formalism is very general, treating fermionic or bosonic systems on equal footing and allowing interactions. Our approach is based on an expansion of the full density operator in orders of the driving speed, and is consistent with the first and second laws of thermodynamics (at least up to the second order in driving speed). In addition, for a model based on system-bath separation, we can formulate the thermodynamics quantities for the sub-system only (assuming that we drive only $\hat H_D$ (and not $\hat H_I$). When applied to the resonant-level model, our results reduce to previous known results.\cite{PhysRevB.93.115318} When electron-electron interactions are included, thermodynamic quantities show interesting Kondo resonances at low temperature. Finally, we emphasize that our current approach can be easily extended to multiple levels and beyond the wide-band limit. Future work must address the outstanding quantities of quantum thermodynamics of a system in the presence of multiple baths under nonequilibrium conditions.

\begin{acknowledgments}

We would like to thank F. von Oppen for helpful discussions. This work was supported by the (U.S.) Air Force Office of Scientific Research (USAFOSR) PECASE award under AFOSR Grant No. FA9950-13-1-0157(JES) and the U.S. National Science Foundation (Grant No. CHE1665291) and the German Research Foundation (DFG TH 820/11-1) (AN).

\end{acknowledgments}

\appendix

\section{Evaluating $\dot W^{(2)}_D$} \label{app:W2}

For the resonant level model, the steps taken to evaluate $\dot W_D^{(2)}$ are very similar to the steps taken in the Supplemental Material of Ref. \citen{PhysRevLett.119.046001}. Similar procedures will also be taken to evaluate $\dot E_D^{(2)}-  \mu \dot N_D^{(2)} $ in Appendix \ref{app:E2}. 

For convenience, we denote the non-interacting Hamiltonian (Eq. \eqref{eq:resonant}) as 
\begin{eqnarray}
\hat H = \sum_{pq} \mathcal{H}_{pq} \hat c^+_p \hat c_q 
\end{eqnarray}
For simplicity, we assume $\mathcal{H}_{pq}$ is real. (For Abe: Indeed, complex number makes thing very complicated. I cannot address the complex number case right now. We always try to avoid complex number case, as also chosen in Ref. \cite{PhysRevLett.119.046001}.)
In the single particle basis, 
\begin{eqnarray}
\sigma_{qp} = {\rm Tr}( \hat c^+_p \hat c_q \hat \rho )  
\end{eqnarray}
At steady state, $\sigma^{(0)}_{qp} = f(\mathcal{H})_{qp}$, 
where $f$ is the Fermi function. Note that for the model in Eq. \eqref{eq:resonant}, only $\epsilon_d(t)$ depends on $t$. 
According to the Markovian approximation,  $\sigma^{(1)}$ can be written as 
\begin{eqnarray}
\sigma^{(1)} = -  \dot \epsilon_d \int_0^{\infty} \exp(- i \mathcal{H} t/\hbar)  \frac{\partial \sigma^{(0)} }{\partial \epsilon_d} \exp(i \mathcal{H} t/\hbar) dt 
\end{eqnarray}

Let the eigen basis of 
$\mathcal{H}$ be denoted by $\{| n \rangle \}$, so that $\mathcal{H} \ket n = \epsilon_n \ket n$. 
For the resonant model in Eq. \eqref{eq:resonant}, we can perform  following manipulation in the single particle basis, 
\begin{eqnarray}
 {\rm Tr}(\frac{d \hat H_D}{dt}  \hat \rho)=  \dot \epsilon_d {\rm Tr}(\frac{\partial \hat H_D}{\partial \epsilon_d}  \hat \rho) =   \dot \epsilon_d  {\rm Tr}( \hat d^+ \hat d  \hat \rho) =  \dot \epsilon_d   \bra d \sigma   \ket d 
\end{eqnarray}
where $\bra d \sigma   \ket d$ is the matrix element for dot level. $\dot W^{(2)}_D$ can then be rewritten as 
\begin{eqnarray}
\dot W^{(2)}_D  &=& \dot \epsilon_d  {\rm Tr}(\frac{\partial \hat H_D}{\partial \epsilon_d}  \hat \rho^{(1)}) =  \dot \epsilon_d  \bra d \sigma^{(1)}  \ket d \nonumber \\
&=& -\dot \epsilon_d^2  \int_0^{\infty} \bra d \exp(- i \mathcal{H} t/\hbar)  \frac{\partial \sigma^{(0)} }{\partial \epsilon_d} \exp(i \mathcal{H} t/\hbar) \ket d dt \nonumber \\
&=& -\dot \epsilon_d^2 \sum_{mn} \int_0^{\infty} \bra d m \rangle \exp(- i \epsilon_m t/\hbar)  \bra m\frac{\partial \sigma^{(0)} }{\partial \epsilon_d}  \ket n \exp(i \epsilon_n t /\hbar ) \langle n \ket d dt \nonumber \\
&=& - \hbar \dot \epsilon_d^2 \sum_{mn} \bra d m \rangle  \bra m\frac{\partial \sigma^{(0)} }{\partial \epsilon_d}  \ket n \langle n \ket d \frac i{\epsilon_m - \epsilon_n+i\eta} 
\end{eqnarray}
Since $\dot W^{(2)}_D$ is real, we can take the real part of the above equation. Note that all matrix elements above are real ($\mathcal{H}_{pq}$ is real), such that 
\begin{eqnarray}
\dot W^{(2)}_D  &=& - \pi \hbar \dot \epsilon_d^2  \sum_{mn} \bra d m \rangle  \bra m\frac{\partial \sigma^{(0)} }{\partial \epsilon_d}  \ket n \langle n \ket d \delta(\epsilon_m - \epsilon_n)
\end{eqnarray}
As shown in Appendix \ref{app:dSigmadR} (and also in the Supplemental Material of Ref. \citen{PhysRevLett.119.046001}), we can evaluate 
\begin{eqnarray}
\bra m\frac{\partial \sigma^{(0)} }{\partial \epsilon_d}  \ket n = \bra m\frac{\partial \mathcal{H} }{\partial \epsilon_d}  \ket n   \frac { f(\epsilon_m)- f(\epsilon_n) } {\epsilon_m - \epsilon_n }  
\end{eqnarray}
For the resonant-level, $\frac{\partial \mathcal{H} }{\partial \epsilon_d}=\ket d \bra d $, Thus, using the identities
\begin{eqnarray}
\delta(\epsilon_m - \epsilon_n)  \frac { f(\epsilon_m)- f(\epsilon_n) } {\epsilon_m - \epsilon_n }=\delta(\epsilon_m - \epsilon_n)  \frac { \partial f(\epsilon_m) } { \partial \epsilon_m  }   
 \end{eqnarray}
and
\begin{eqnarray}
 \delta(\epsilon_m - \epsilon_n) = \int d\epsilon \delta(\epsilon - \epsilon_n)\delta(\epsilon - \epsilon_m) , 
 \end{eqnarray}
we can proceed to simplify 
\begin{eqnarray}
\dot W^{(2)}_D  
&=&- \pi \hbar \dot \epsilon_d^2  \sum_{mn} \bra d m \rangle  \bra m\frac{\partial  \mathcal{H} }{\partial \epsilon_d}  \ket n \langle n \ket d \delta(\epsilon_m - \epsilon_n)  \frac { f(\epsilon_m)- f(\epsilon_n) } {\epsilon_m - \epsilon_n }\nonumber \\
&=&- \pi \hbar \dot \epsilon_d^2\sum_{mn} \bra d m \rangle  \bra m\frac{\partial  \mathcal{H} }{\partial \epsilon_d}  \ket n \langle n \ket d \delta(\epsilon_m - \epsilon_n)  \frac { \partial f(\epsilon_m) } { \partial \epsilon_m  } \nonumber \\
&=&- \pi \hbar \dot \epsilon_d^2  \sum_{mn} \bra d m \rangle  \bra m d\rangle \langle d \ket n \langle n \ket d \delta(\epsilon_m - \epsilon_n)  \frac { \partial f(\epsilon_m) } { \partial \epsilon_m  } \nonumber \\
&=& - \pi \hbar \dot \epsilon_d^2  \sum_{mn}  \int d\epsilon \bra d m \rangle \delta(\epsilon - \epsilon_m)  \bra m d\rangle \langle d \ket n \delta(\epsilon - \epsilon_n) \langle n \ket d   \frac { \partial f(\epsilon) } { \partial \epsilon  } \nonumber \\
&=& - \hbar \dot \epsilon_d^2    \int \frac{d\epsilon}{\pi} \bra d \mbox{Im} G  \ket d \bra d \mbox{Im} G\ket d   \frac { \partial f(\epsilon) } { \partial \epsilon  }
\end{eqnarray}
Here $G$ is the single particle Green's function
\begin{eqnarray}
G= \sum_{m} \ket m \frac{1}{\epsilon-\epsilon_m + i\eta} \bra m 
\end{eqnarray}  
and the imaginary part is 
\begin{eqnarray}
\mbox{Im} G= -\pi \sum_{m} \ket m \delta( \epsilon-\epsilon_m ) \bra m  
\end{eqnarray}
The dot level matrix element  of  $\mbox{Im} G$ gives the spectral function $A= -2 \bra d \mbox{Im} G  \ket d $, such that 
\begin{eqnarray} \label{eq:appW2_sub}
\dot W^{(2)}_D  
= - \frac{\hbar \dot \epsilon_d^2}2   \int \frac{d\epsilon}{2\pi} A^2   \frac { \partial f(\epsilon) } { \partial \epsilon  }
\end{eqnarray}

\section{Evaluating $\dot E^{(2)}_D-  \mu \dot N^{(2)}_D$} \label{app:E2}

Similarly, we can now evaluate $\dot E^{(2)}_D -  \mu \dot N^{(2)}_D$. Note that 
\begin{eqnarray}
 {\rm Tr}(\hat N_D \hat \rho) =  {\rm Tr}(  \hat d^+ \hat d  \hat \rho) = \bra d   \sigma   \ket d 
\end{eqnarray}
and 
\begin{eqnarray}
 {\rm Tr}(\hat H_D^{eff} \hat \rho) =  {\rm Tr}( (\epsilon_d \hat d^+ \hat d  + \frac12 \sum_k V_k (\hat c_k^+ \hat d + \hat d^+ \hat c_k) )\hat \rho) \nonumber \\
= \Re {\rm Tr}( (\epsilon_d \hat d^+ \hat d  +  \sum_k V_k  \hat d^+ \hat c_k) )\hat \rho) = \Re \bra d \mathcal{H}   \sigma   \ket d 
\end{eqnarray}
Therefore, 
\begin{eqnarray}
\dot E^{(2)}_D -  \mu \dot N^{(2)}_D 
&=& \dot \epsilon_d\frac{\partial  }{\partial \epsilon_d} {\rm Tr}( (\hat H_D - \mu \hat N_D ) \hat \rho^{(1)} )  \nonumber \\
&=&  \dot \epsilon_d\frac{\partial  }{\partial \epsilon_d} \Re \bra d (\mathcal{H} - \mu ) \sigma^{(1)}   \ket d \nonumber \\
&=&\dot \epsilon_d\frac{\partial  }{\partial \epsilon_d} \Re \sum_{mn} \bra d (\mathcal{H} - \mu ) \ket m \bra m \sigma^{(1)} \ket n \langle n  \ket d \nonumber \\
&=& \dot \epsilon_d \frac{\partial  }{\partial \epsilon_d} \Re \sum_{mn} (\epsilon_m - \mu) \bra d  m \rangle \bra m \sigma^{(1)} \ket n \langle n  \ket d 
\end{eqnarray}
Again, all matrix elements are real. If we look at the following matrix element, using the results above, we arrive at
\begin{eqnarray}
\Re \bra m \sigma^{(1)} \ket n &=& - \dot \epsilon_d \Re \int_0^{\infty} \bra m \exp(- i \mathcal{H} t/\hbar)  \frac{\partial \sigma^{(0)} }{\partial \epsilon_d} \exp(i \mathcal{H} t/\hbar) \ket n dt \nonumber \\
 &=& -\pi\hbar \dot \epsilon_d \bra m\frac{\partial \sigma^{(0)} }{\partial \epsilon_d}  \ket n \delta(\epsilon_m - \epsilon_n)  \nonumber \\
 &=& -\pi\hbar \dot \epsilon_d \bra m\frac{\partial  \mathcal{H} }{\partial \epsilon_d}  \ket n  \frac { f(\epsilon_m)- f(\epsilon_n) } {\epsilon_m - \epsilon_n } \delta(\epsilon_m - \epsilon_n)  \nonumber \\
 &=& - \pi\hbar\dot \epsilon_d \bra m\frac{\partial  \mathcal{H} }{\partial \epsilon_d}  \ket n  \frac {\partial f(\epsilon_m)} {\partial \epsilon_m  } \delta(\epsilon_m - \epsilon_n)  \nonumber \\
  &=& - \pi\hbar \dot \epsilon_d  \bra m  d \rangle \langle d \ket n  \frac {\partial f(\epsilon_m)} {\partial \epsilon_m  } \delta(\epsilon_m - \epsilon_n)  
\end{eqnarray}
Accordingly,  we can evaluate 
\begin{eqnarray}
\dot E^{(2)}_D -  \mu \dot N^{(2)}_D &=& -\pi\hbar \dot \epsilon_d^2 \frac{\partial  }{\partial \epsilon_d} \sum_{mn} (\epsilon_m - \mu) \bra d  m \rangle \bra m    d \rangle \langle d  \ket n \langle n  \ket d \frac {\partial f(\epsilon_m)} {\partial \epsilon_m  } \delta(\epsilon_m - \epsilon_n) \nonumber \\
&=&- \pi\hbar \dot \epsilon_d^2  \frac{\partial  }{\partial \epsilon_d} \sum_{mn} \int d\epsilon (\epsilon - \mu) \bra d  m \rangle \delta(\epsilon - \epsilon_m)  \bra m    d \rangle \langle d  \ket n \delta(\epsilon - \epsilon_n) \langle n  \ket d \frac {\partial f(\epsilon)} {\partial \epsilon  }  \nonumber \\
&=&-\hbar\dot \epsilon_d^2  \frac{\partial  }{\partial \epsilon_d}  \int \frac{d\epsilon}{\pi} (\epsilon - \mu) \bra d  \mbox{Im} G    \ket d  \bra d  \mbox{Im} G \ket d \frac {\partial f(\epsilon)} {\partial \epsilon  }  \nonumber \\
&=&-\frac{\hbar \dot \epsilon_d^2}2  \frac{\partial }{\partial \epsilon_d}  \int \frac{d\epsilon}{2\pi} (\epsilon - \mu) A^2 \frac {\partial f(\epsilon)} {\partial \epsilon  } \nonumber   \\
&=&-\frac{\hbar \dot \epsilon_d^2}2   \int \frac{d\epsilon}{2\pi} (\epsilon - \mu) \frac{\partial A^2 }{\partial \epsilon_d } \frac {\partial f(\epsilon)} {\partial \epsilon  }  
\end{eqnarray}

Up until  to now, our results have not relied on the wide-band approximation. 
In the wide-band limit, $A= \frac{\Gamma}{(\epsilon-\epsilon_d)^2 + (\Gamma/2)^2}$, such that $  \frac{\partial  }{\partial \epsilon_d } A = -  \frac{\partial  }{\partial \epsilon } A $. If we integrate by parts, we find 
\begin{eqnarray} \label{eq:E2_subapp}
\dot E^{(2)}_D  -  \mu \dot N^{(2)}_D 
&=& \frac{\hbar \dot \epsilon_d^2}2   \int \frac{d\epsilon}{2\pi}  (\epsilon - \mu)  \frac{\partial A^2 }{\partial \epsilon } \frac {\partial f(\epsilon)} {\partial \epsilon  }  \nonumber \\
&=&-\frac{\hbar \dot \epsilon_d^2}2   \int \frac{d\epsilon}{2\pi}   A^2 \frac{\partial }{\partial \epsilon } ( (\epsilon - \mu)  \frac {\partial f(\epsilon)} {\partial \epsilon  } )  \nonumber \\
&=&-\frac{\hbar \dot \epsilon_d^2}2  \int \frac{d\epsilon}{2\pi}   A^2 ( (\epsilon - \mu)  \frac {\partial^2 f(\epsilon)} {\partial \epsilon^2  }  +  \frac{\partial f}{\partial \epsilon } ) 
\end{eqnarray}
From Eq. \eqref{eq:E2_subapp} and Eq. \eqref{eq:appW2_sub}, we recover the results in Ref. \citen{PhysRevB.94.035420}: 
\begin{eqnarray}
\dot Q^{(2)}_D  =   \dot E^{(2)}_D - \mu \dot N^{(2)}_D - \dot W^{(2)}_D = - \frac{\hbar \dot \epsilon_d^2}2    \int d\epsilon   (\epsilon - \mu) A^2   \frac {\partial^2 f(\epsilon)} {\partial \epsilon^2  }   
\end{eqnarray}

\section{Evaluating $\bra m\frac{\partial \sigma^{(0)} }{\partial \epsilon_d}  \ket n $} \label{app:dSigmadR}
We note that 
\begin{eqnarray}
\frac{\partial}{\partial \epsilon_d}    f(\epsilon_m) \delta_{mn} &=& \frac{\partial}{\partial \epsilon_d}  \bra m  \sigma^{(0)}  \ket n =\bra m\frac{\partial \sigma^{(0)} }{\partial \epsilon_d}  \ket n +  \frac{\partial \bra m }{\partial \epsilon_d}   \sigma^{(0)}  \ket n +  \bra m    \sigma^{(0)} \frac{\partial \ket n }{\partial \epsilon_d}  \nonumber  \\
&=& \bra m\frac{\partial \sigma^{(0)} }{\partial \epsilon_d}  \ket n +  f(\epsilon_n) \frac{\partial \bra m }{\partial \epsilon_d}   \ket n + f(\epsilon_m) \bra m  \frac{\partial \ket n }{\partial \epsilon_d}  \nonumber  \\
&=& \bra m\frac{\partial \sigma^{(0)} }{\partial \epsilon_d}  \ket n +  ( f(\epsilon_m)- f(\epsilon_n) ) \bra m  \frac{\partial \ket n }{\partial \epsilon_d}  
\end{eqnarray}
Similarly, 
\begin{eqnarray}
\frac{\partial}{\partial \epsilon_d}   \epsilon_m  \delta_{mn} = \frac{\partial}{\partial \epsilon_d}  \bra m  \mathcal{H}  \ket n 
= \bra m\frac{\partial \mathcal{H} }{\partial \epsilon_d}  \ket n +  ( \epsilon_m - \epsilon_n) \bra m  \frac{\partial \ket n }{\partial \epsilon_d}  
\end{eqnarray}

At this point, we multiply  $\frac { f(\epsilon_m)- f(\epsilon_n) } {\epsilon_m - \epsilon_n } $ on both sides of the above equation:  
\begin{eqnarray}
\bra m\frac{\partial \mathcal{H} }{\partial \epsilon_d}  \ket n   \frac { f(\epsilon_m)- f(\epsilon_n) } {\epsilon_m - \epsilon_n }   = \frac { f(\epsilon_m)- f(\epsilon_n) } {\epsilon_m - \epsilon_n } \frac{\partial}{\partial \epsilon_d}   \epsilon_m  \delta_{mn} - ( f(\epsilon_m)- f(\epsilon_n) ) \bra m  \frac{\partial \ket n }{\partial \epsilon_d}  \nonumber \\
\end{eqnarray}
Note that 
\begin{eqnarray}
 \frac { f(\epsilon_m)- f(\epsilon_n) } {\epsilon_m - \epsilon_n } \frac{\partial}{\partial \epsilon_d}   \epsilon_m  \delta_{mn} 
= \frac{\partial}{\partial \epsilon_d}    f(\epsilon_m) \delta_{mn}, 
\end{eqnarray}
and therefore, 
\begin{eqnarray}
\bra m\frac{\partial \mathcal{H} }{\partial \epsilon_d}  \ket n   \frac { f(\epsilon_m)- f(\epsilon_n) } {\epsilon_m - \epsilon_n }   = \frac{\partial}{\partial \epsilon_d}    f(\epsilon_m) \delta_{mn} - ( f(\epsilon_m)- f(\epsilon_n) ) \bra m  \frac{\partial \ket n }{\partial \epsilon_d} 
=\bra m\frac{\partial \rho_0 }{\partial \epsilon_d}  \ket n  \nonumber \\
\end{eqnarray}

\section{NRG and MFT calculation for the Anderson model} \label{app:NRG}
In a NRG calculation, the Anderson model is mapped onto a semi-infinite chain, 
\begin{eqnarray}
\hat H = \epsilon_d (t) \sum_\sigma \hat d^+_\sigma \hat d_\sigma + U \hat d^+_\uparrow \hat d_\uparrow \hat d^+_\downarrow \hat d_\downarrow+ \sqrt{\frac{\Gamma}{\pi} }\sum_{\sigma} (\hat d^+_\sigma \hat f_{0\sigma}  + \hat f_{0\sigma}^+ \hat d_\sigma) \nonumber \\
+ \sum_{n\sigma} t_{n}  ( \hat f_{n\sigma}^+ \hat f_{n+1\sigma}  +  \hat f_{n+1\sigma}^+ \hat f_{n\sigma} ) 
\end{eqnarray}
where $t_n$ decays exponentially with $n$ (the exact form of $t_n$ are given in Ref. \citen{nrgreview}). 
Furthermore,  the subsystem $\hat H_D$ is  
\begin{eqnarray}
\hat H_D = \epsilon_d (t) \sum_\sigma \hat d^+_\sigma \hat d_\sigma + U \hat d^+_\uparrow \hat d_\uparrow \hat d^+_\downarrow \hat d_\downarrow+ \frac12 \sqrt{\frac{\Gamma}{\pi} }\sum_{\sigma} (\hat d^+_\sigma \hat f_{0\sigma}  + \hat f_{0\sigma}^+ \hat d_\sigma) 
\end{eqnarray}
Using eigenstates of $\hat H$ from NRG, $\hat H | \Psi_I \rangle = E_I | \Psi_I \rangle  $, just as was shown in the supplemental material of Ref. \citen{PhysRevLett.119.046001}, 
 we can now calculate the first order energy as follows (with NRG): 
\begin{eqnarray}
E^{(1)}_D&=&{\rm Tr}( \hat H_D  \hat \rho^{(1)} ) 
= - \dot x \int_0^\infty  {\rm Tr}( \hat H_D e^{-i \hat H t'/\hbar} \frac{\partial}{\partial x} \hat \rho^{(0)} e^{i \hat  H t'/\hbar} )  dt' \\
&=& \dot x \frac {\pi\hbar\beta} 2 \frac{\partial \epsilon_d }{\partial x} \sum_{IJ}  \langle \Psi_I | \hat H_D | \Psi_J \rangle   \langle \Psi_J |  \delta \hat n | \Psi_I \rangle   \frac{ e^{-\beta E_J} + e^{-\beta E_I} }{Z}  \delta( E_J - E_I) 
\end{eqnarray}
where $\delta \hat n =  \sum_\sigma \left( \hat d^+_\sigma \hat d_\sigma -  \frac1Z{\sum_I \langle \Psi_I |  \hat d^+_\sigma \hat d_\sigma \Psi_I \rangle e^{-\beta E_I} }  \right) $, and $Z= \sum_J e^{-\beta E_J}$. Further details of the actual NRG calculation can be found in the supplemental material of Ref. \citen{PhysRevLett.119.046001}. For the NRG calculations in Sec. \ref{sec:sub-system}, we set the logarithmic discretization parameter to $\Lambda=2$. At each step, we keep up to 500 states. 

Finally, let us discuss MFT. 
When treated within a mean-field level, the total Hamiltonian becomes quadratic:
\begin{eqnarray}
\hat H_{MFT} = E_{eff} \sum_\sigma \hat d^+_\sigma \hat d_\sigma + \sum_{k\sigma} V_k ( \hat d^+_\sigma \hat c_{k\sigma}  + \hat c_{k\sigma}^+ \hat d_\sigma) + \sum_{k\sigma} \epsilon_{k} \hat c_{k\sigma}^+ \hat c_{k\sigma}  
\end{eqnarray}
Let us assume a spin restricted solution so that $n_\uparrow=n_\downarrow$, and
\begin{eqnarray} \label{eq:EeffMFT}
E_{eff} = \epsilon_d + n_\uparrow U
\end{eqnarray}
where  
\begin{eqnarray} \label{eq:n_up}
n_\uparrow = \int \frac{d\epsilon}{2\pi}  \frac{\Gamma} { (\epsilon - E_{eff})^2 + (\Gamma/2)^2 } f(\epsilon) 
\end{eqnarray}
Eq. \eqref{eq:EeffMFT} and Eq. \eqref{eq:n_up} have to be solved self consistently. With $E_{eff}$, the MFT solution of $E^{(1)}_D $ is 
\begin{eqnarray}
E^{(1)}_D  = - 2\times \dot x \frac{\hbar} 2   \frac{\partial E_{eff} }{\partial x}   \int \frac{d\epsilon}{2\pi} \epsilon \left( \frac{\Gamma} { (\epsilon - E_{eff})^2 + (\Gamma/2)^2 } \right)^2 \frac {\partial f(\epsilon)} {\partial \epsilon  }  
\end{eqnarray}
The factor 2 in front of the above equation counts for spin degeneracy. 

{ 
\section{Proof of Eq. \eqref{eq:Sad}} \label{app:SSS}
Here we calculate the first order correction to $S^{(0)}$. 
To explicitly indicate the small parameter, we write 
\begin{eqnarray}
\hat \rho = \hat \rho^{(0)} + \lambda \hat \rho^{(1)} + \lambda^2 \hat \rho^{(2)} + \cdots
\end{eqnarray}
where we have used the power of $\lambda$ to indicate the order of small parameters. Our goal is to expand $S$ 
\begin{eqnarray}
S = -k_B {\rm Tr} ( \hat \rho \ln \hat \rho ) = S^{(0)} + \lambda S^{(1)} +  \lambda^2 S^{(2)} + \cdots
\end{eqnarray} 
The zeroth order then can be written as 
\begin{eqnarray}
S^{(0)}  = -k_B {\rm Tr} ( \hat \rho \ln \hat \rho )|_{\lambda = 0} = -k_B {\rm Tr} ( \hat \rho^{(0)}  \ln \hat \rho^{(0)} ) 
\end{eqnarray}
The first order correction is 
\begin{eqnarray}
S^{(1)} = -k_B \frac{d}{d\lambda} {\rm Tr} ( \hat \rho \ln \hat \rho )|_{\lambda = 0} =  -k_B  {\rm Tr} ( \frac{d \hat \rho}{d\lambda}  \ln \hat \rho )|_{\lambda = 0} -k_B {\rm Tr} ( \hat \rho \frac{d }{d\lambda}  \ln \hat \rho )|_{\lambda = 0} \nonumber \\
= -k_B  {\rm Tr} ( \hat \rho^{(1)}  \ln \hat \rho^{(0)} )  - k_B {\rm Tr} ( \hat \rho  \hat \rho^{-1}  \frac{d }{d\lambda} \hat \rho )|_{\lambda = 0} = -k_B  {\rm Tr} ( \hat \rho^{(1)}  \ln \hat \rho^{(0)} )  
\end{eqnarray}
Here, 
we have used that  $ {\rm Tr} ( \hat \rho \frac{d }{d\lambda}  \ln \hat \rho ) =   {\rm Tr} ( \hat \rho  \hat \rho^{-1} \frac{d }{d\lambda}\hat \rho )  $. To prove this identity,} we assume that $\ln \hat \rho$ can be formally expanded in a power series
 $\ln \hat \rho  = \sum_{n=1}^\infty (-1)^{n+1} (\hat \rho - \hat 1 )^n / n$. Using the cyclic property of the trace, this leads to
 \begin{eqnarray}
{\rm Tr} ( \hat \rho \frac{d }{d\lambda}  \ln \hat \rho )  = {\rm Tr} ( \hat \rho  \sum_{n=1}^\infty (-1)^{n+1} (\hat \rho - \hat 1 )^{n-1}   \frac{d }{d\lambda} \hat \rho ) = {\rm Tr} ( \hat \rho \hat \rho^{-1}  \frac{d }{d\lambda} \hat \rho )
\end{eqnarray}
where we have used the formal expansion of $\hat \rho^{-1} = ( \hat 1 + \hat \rho - \hat 1)^{-1}$. Admittedly, in this derivation we have used formal expansions that numerically converge only when $||\hat \rho||$ is close enough to 1, which is not necessarily true here. Note, however, that we have used this expansion only to prove the identity  $ {\rm Tr} ( \hat \rho \frac{d }{d\lambda}  \ln \hat \rho ) =   {\rm Tr} ( \hat \rho  \hat \rho^{-1} \frac{d }{d\lambda}\hat \rho )  $,   where the ${\rm Tr}$ helps to remove the ordering ambiguity between $\hat \rho$  (or $\hat \rho^{-1}$) and $d\hat \rho/d\lambda$. Alternatively we could prove the same identity using (again, formally) the representation of  $\hat \rho$.


\begin{thebibliography}{62}
\bibitem [{the()}]{thermodynamics:Hs}%
  \BibitemOpen
  \href@noop {} {}\bibinfo {note} {It should be noted however that while Eqs.
  \ref{eq:E_Dnon1}-\ref{eq:E_Dnon2} provide a rigorous definition of energies
  in the D and B subsystems, there is no guarantee that the B part, $\langle
  \hat H \rangle_B$, does not depend on $\epsilon_d$. It can be shown
  \cite{PhysRevB.94.035420} that in the wide band limit, $\langle \hat H
  \rangle_B$ does not depend on $\epsilon_d$. However, in the case of a
  non-Markovian bath, it is found that no splitting of the full Hamiltonian
  into ``dot" and ``bath", $\langle \hat H \rangle = \langle \hat H \rangle_B +
  \langle \hat H\rangle_D$ parts can render all the $\epsilon_d$ dependence of
  $\langle \hat H \rangle$ in the $\langle \hat H \rangle_D $ part, namely, in
  general, $\frac{\partial}{\partial \epsilon_d}\langle \hat H \rangle \ne
  \frac{\partial}{\partial \epsilon_d}\langle \hat H \rangle_D$}\BibitemShut
  {NoStop}%
\end{thebibliography}
\end{document}